\PassOptionsToPackage{table,xcdraw,dvipsnames}{xcolor}
\documentclass[10pt,sigconf,letterpaper,nonacm]{acmart}
\usepackage[T1]{fontenc}
\usepackage[toc,page]{appendix}
\usepackage{graphicx}
\usepackage{color}
\usepackage[utf8]{inputenc}
\usepackage{fancyvrb}
\usepackage{booktabs} 
\usepackage{listings} 
\usepackage{verbatim}
\usepackage{array}
\usepackage{caption}
\usepackage{subcaption}
\usepackage[normalem]{ulem}
\usepackage{float}
\usepackage{tabularray}
\usepackage{longtable}
\usepackage{lscape}
\usepackage{multirow}
\usepackage[table,xcdraw]{xcolor}
\usepackage{url}
\usepackage{cleveref}

\renewcommand\footnotetextcopyrightpermission[1]{} 
\newif\ifshort
\shortfalse

\iftrue
\newcommand{\alisha}[1]{\textcolor{blue}{\noindent[Alisha: #1]}}
\else
\newcommand{\alisha}[1]{}
\fi

\iftrue
\newcommand{\kath}[1]{\textcolor{purple}{\noindent[Katherine: #1]}}
\else
\newcommand{\kath}[1]{}
\fi

\usepackage{xspace}
\newcommand{\violator}{CitM\xspace}
\newcommand{\violators}{CitMs\xspace}
\newcommand{\Violator}{CitM\xspace}
\newcommand{\Violators}{CitMs\xspace}

\usepackage{adjustbox}
\usepackage{array}

\newcolumntype{R}[2]{%
    >{\adjustbox{angle=#1,lap=\width-(#2)}\bgroup}%
    l%
    <{\egroup}%
}

\newcommand{\definitionsfigure}{
    \begin{figure}[t]
        \centering
        \includegraphics[width=\columnwidth]{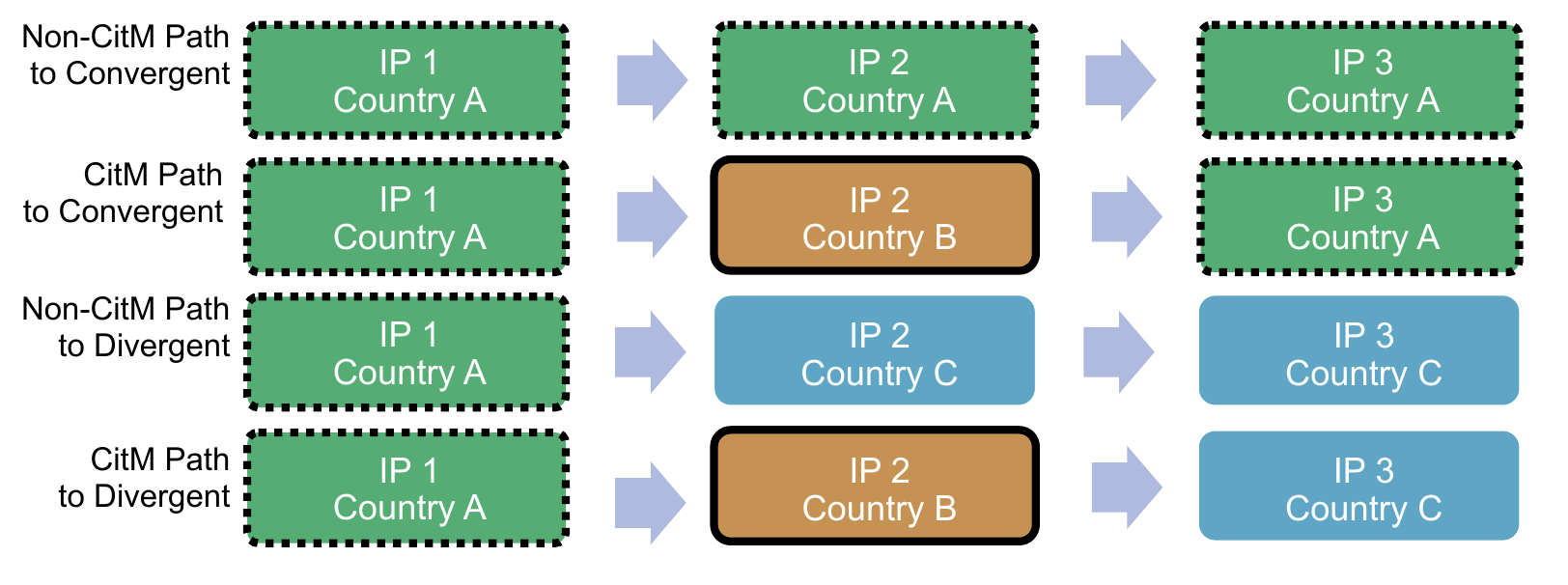}
        \caption{A visualization of our taxonomy. \textnormal{Colors and border patterns indicate the country hosting the client, website, or intermediate hop(s).}}
    \label{fig:definitions}
        \vskip -1em
    \end{figure}
}

\newcommand{\methodology}{
    \begin{figure}[t]
        \centering
        \includegraphics[width=\columnwidth]{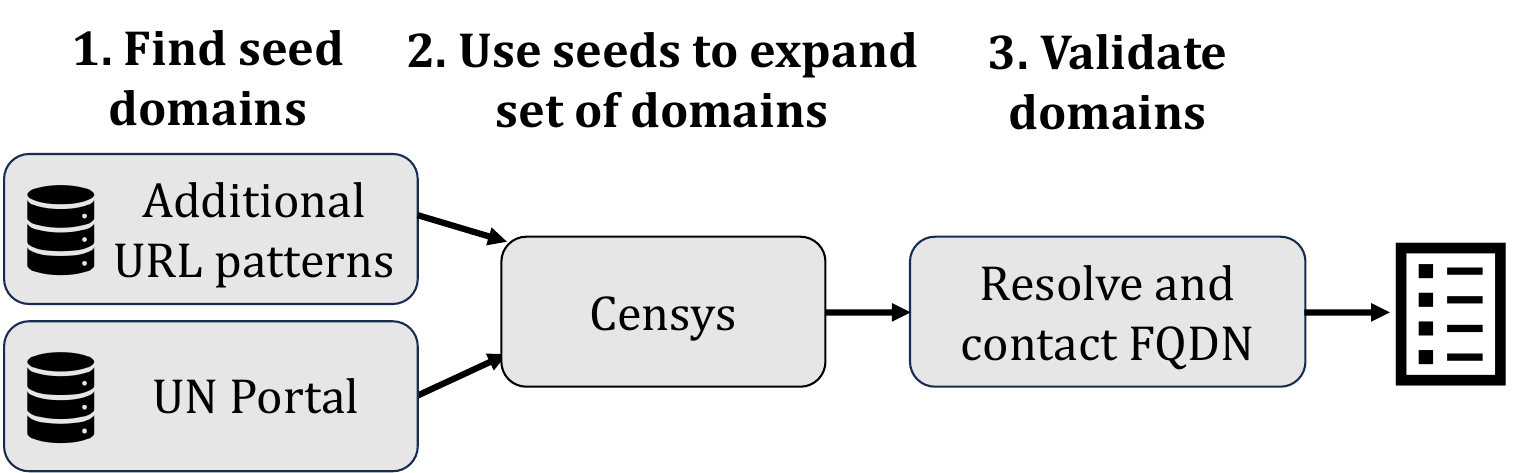}
        \caption{Overview of government domain collection.}
    \label{fig:methodology}
        \vskip -1em
    \end{figure}
}

\DeclareCaptionFormat{oldbanjo}{\textbf{Figure \arabic{figure} (left):} #3}
\newcommand{\BanjoGraphOld}{
    \begin{figure}
        \centering
        \includegraphics[width=\columnwidth]{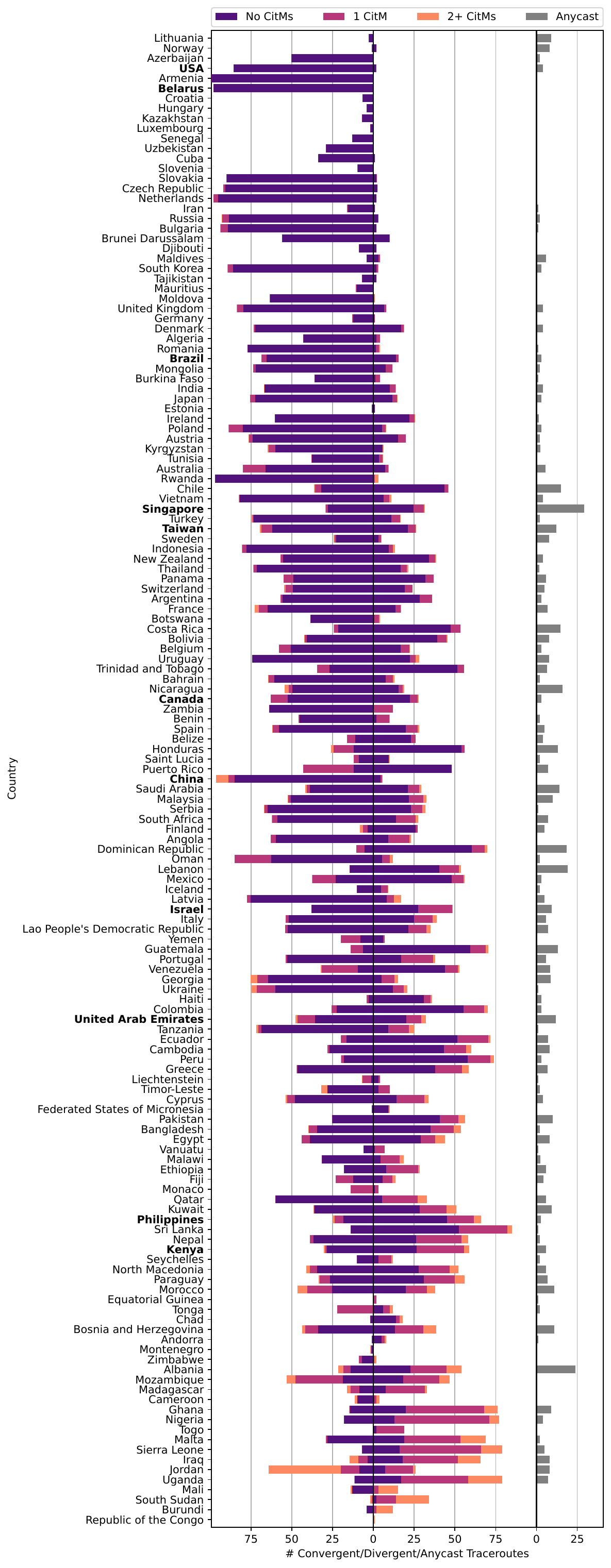}
    \end{figure}
    \begin{figure}[t]
        \captionsetup{format=oldbanjo}
        \caption{Frequency of \violators for each country in our pilot study. Traceroutes are normalized by the number of vantage points. Bolded countries indicate countries of focus in our subsequent study.}
        \label{fig:banjo_old}
    \end{figure}
}

\newcommand{\banjograph}{
    \begin{figure}[t]
        \centering
        \includegraphics[width=\columnwidth]{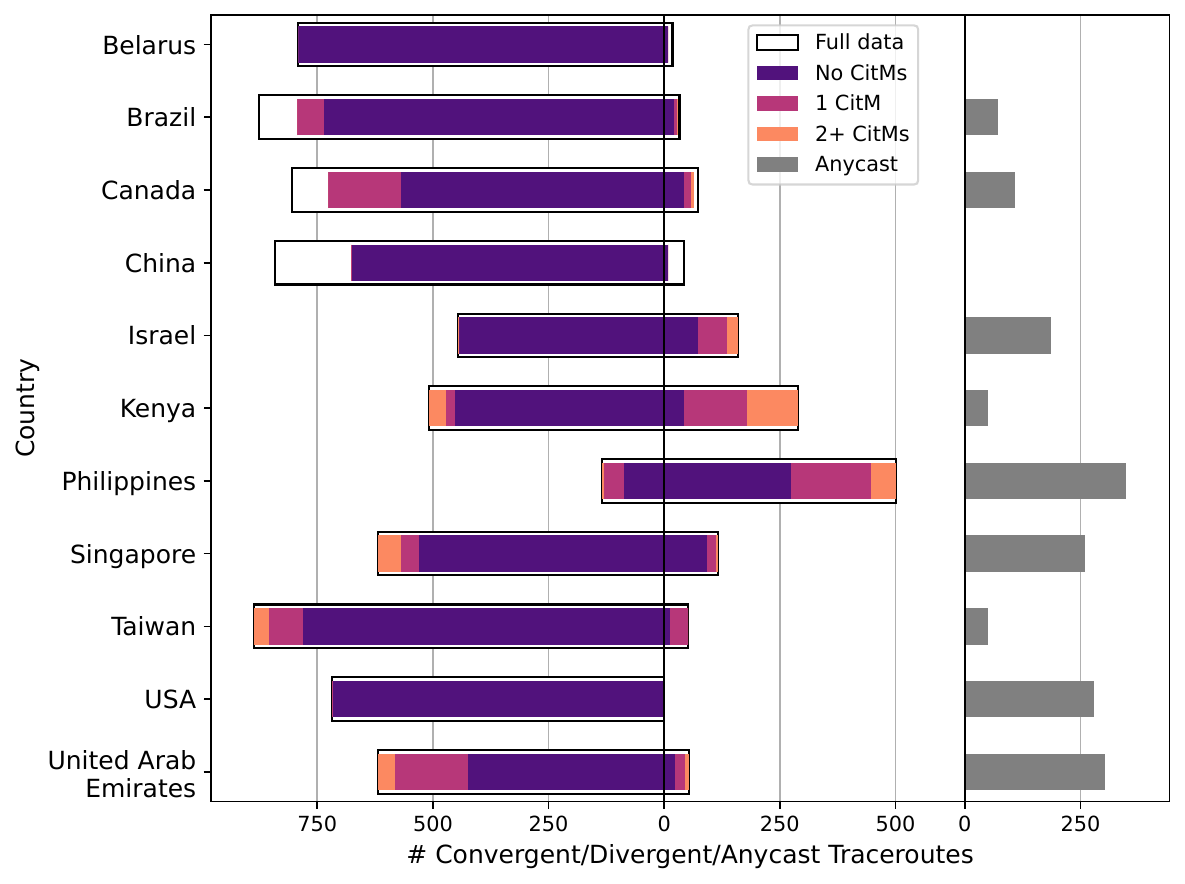}
        \caption{Frequency of \violators in each country's results after data sanitization. Enclosing rectangles show the number of paths for each country before sanitization.}
        \label{fig:banjo}
    \end{figure}
}

\newcommand{\heatmapviolators}{
    \begin{figure*}[t]
        \centering
        \includegraphics[width=0.85\textwidth]{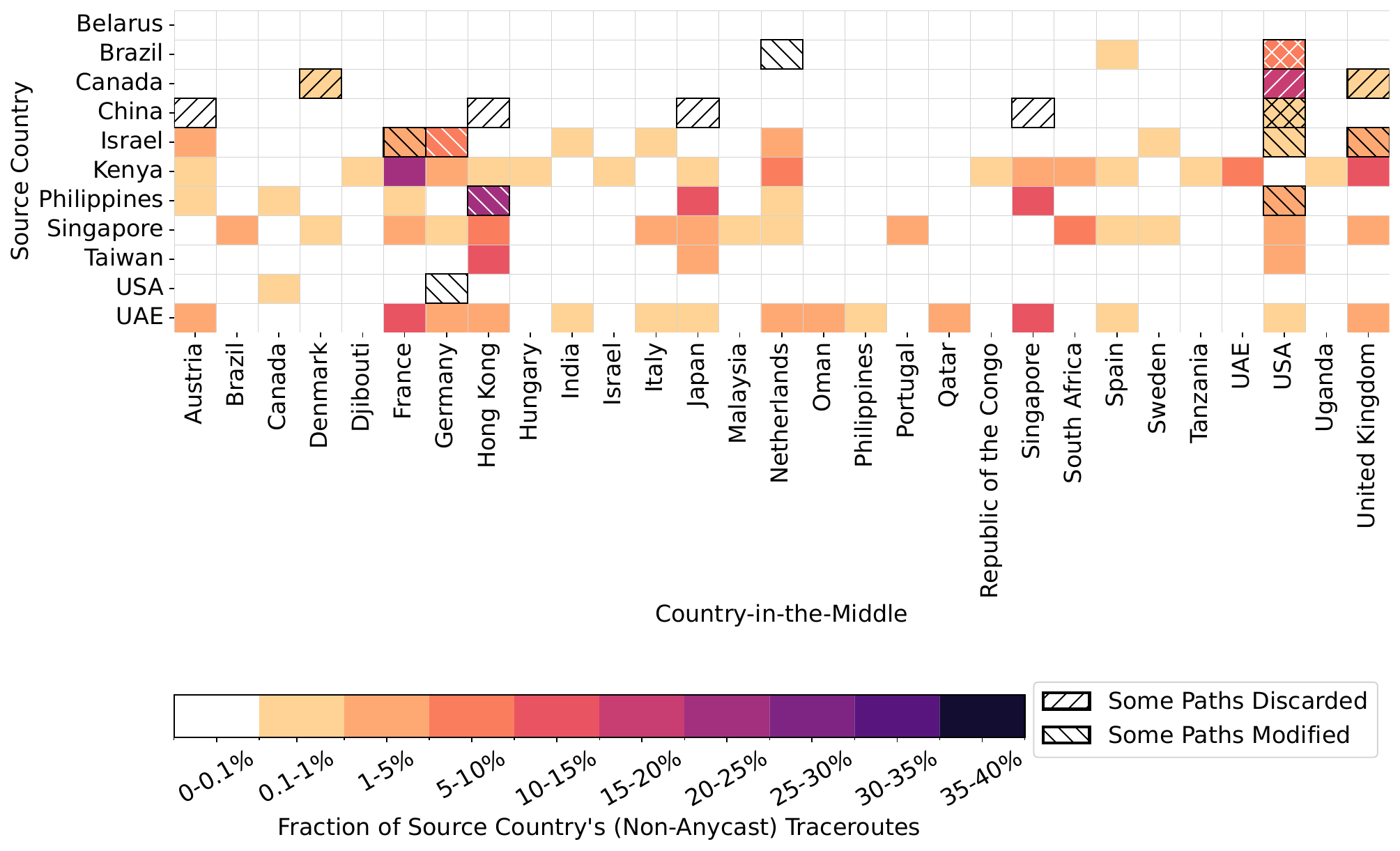}
        \caption{Heatmap of the \violators for convergent and divergent targets after data sanitization. Hatched cells indicate traceroutes being discarded or modified from the full dataset (Section~\ref{sec:navigating}).}
    \label{fig:heatmap_violators}
    \end{figure*}
}

\newcommand{\heatmapconvergentviolators}{
    \begin{figure}[t]
        \begin{subfigure}{\columnwidth}
            \centering
            \includegraphics[width=\columnwidth]{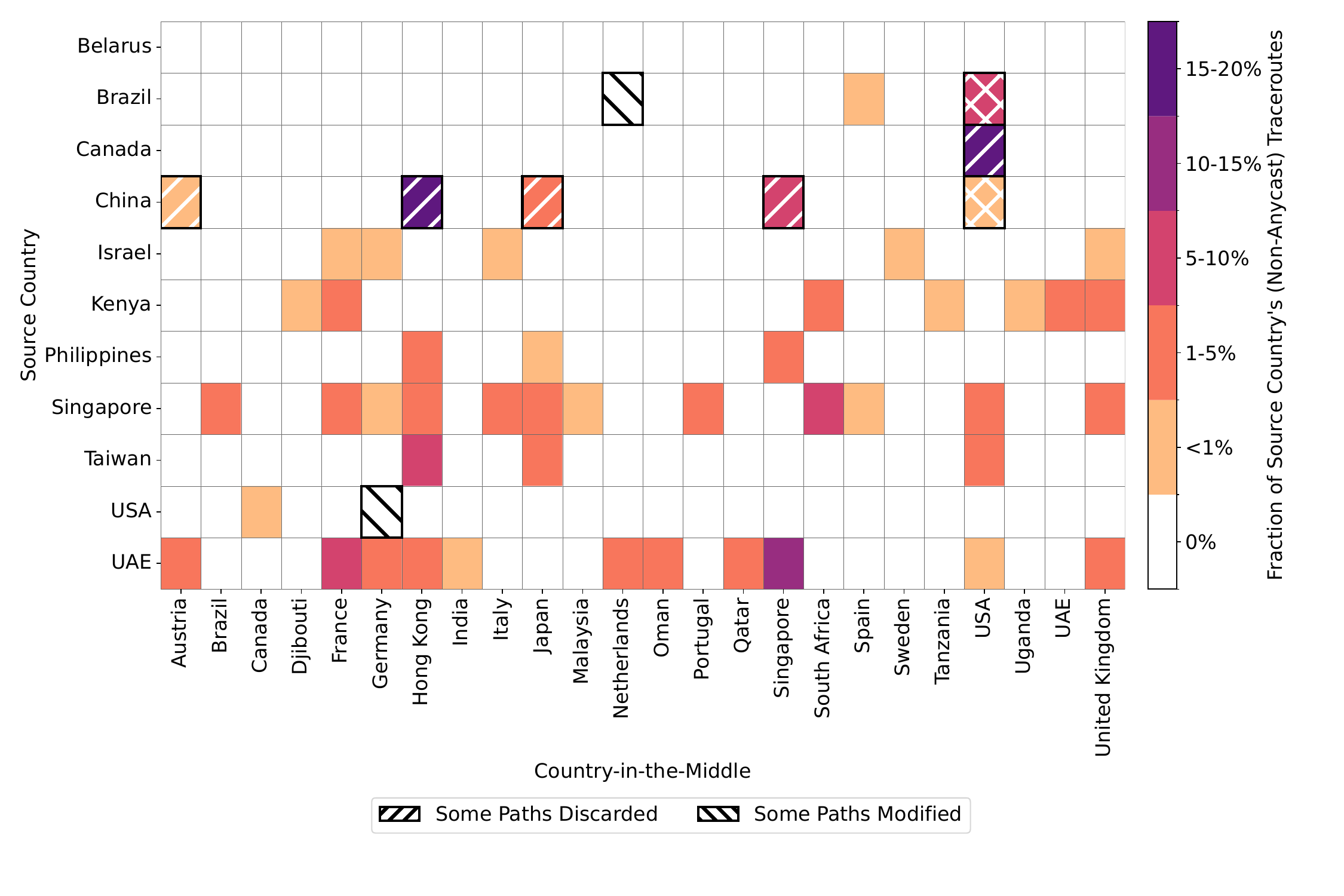}
            \caption{Full (unrefined) dataset}
        \end{subfigure}
        \begin{subfigure}{\columnwidth}
            \centering
            \includegraphics[width=\columnwidth]{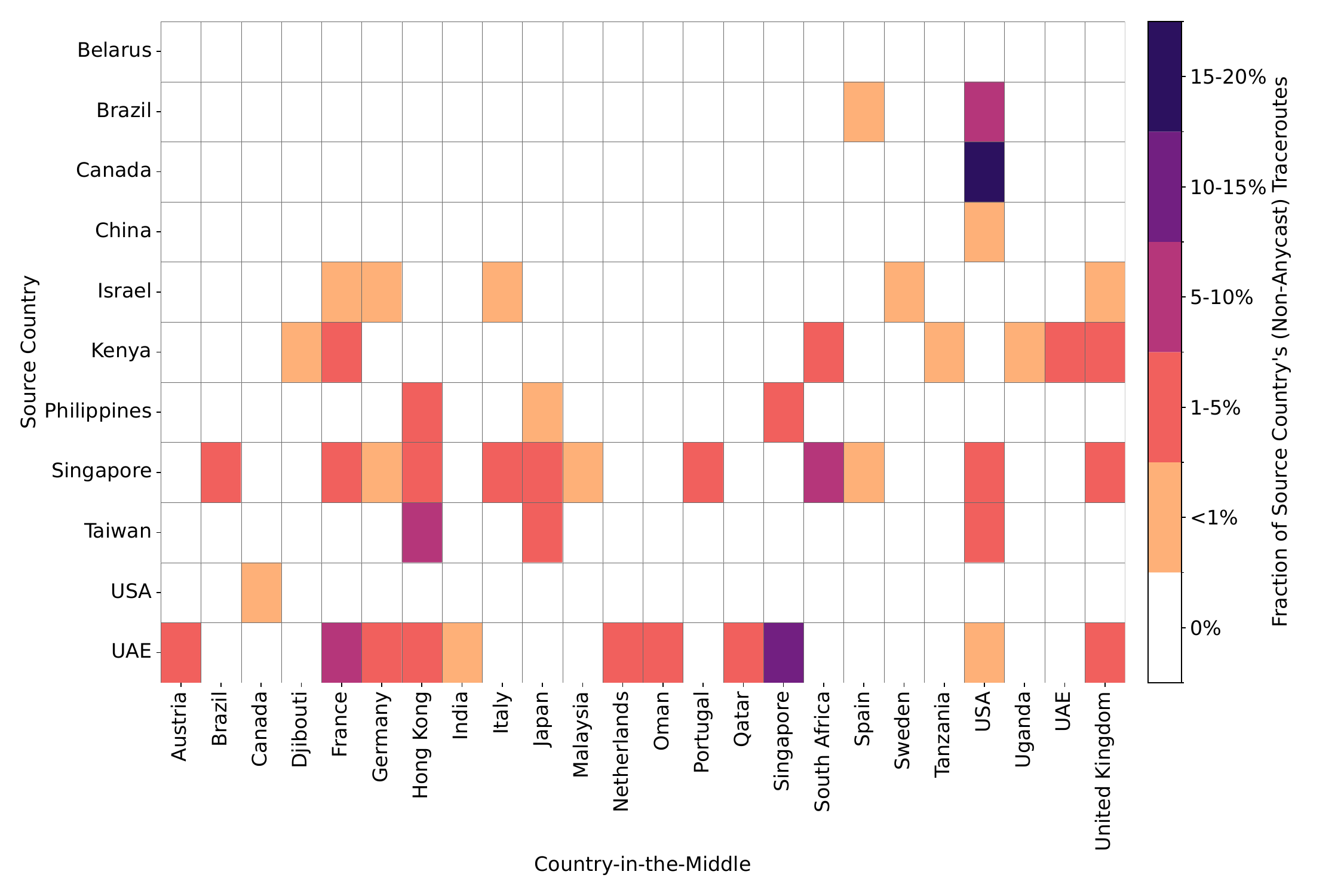}
            \caption{Sanitized dataset}
            \label{fig:heatmap_con_viol_good}
        \end{subfigure}
        \caption{Heatmap of \violators on convergent paths (a) full and (b) sanitized datasets.  Cross-hatching indicates traceroutes that were either discarded or modified from the full dataset.}
        \label{fig:heatmap_convergent_violators}
    \end{figure}
}

\newcommand{\unlabeledcdf}{
    \begin{figure}[t!]
        \centering
        \includegraphics[width=\columnwidth]{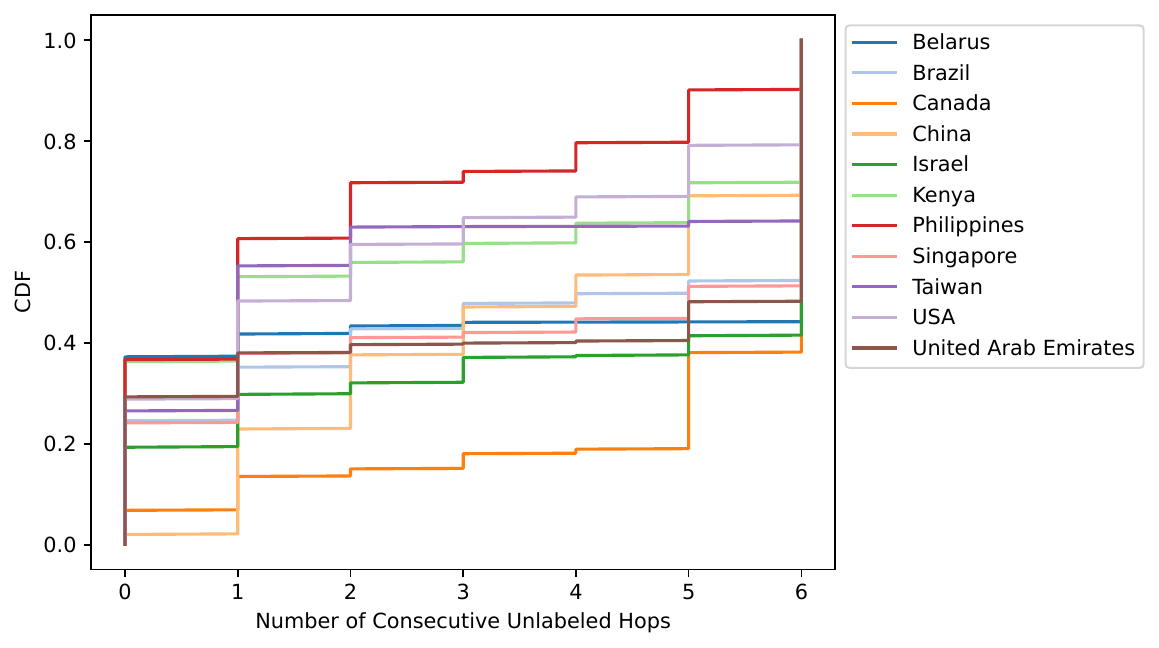}
        \caption{The CDF of the maximum number of consecutive unlabeled hops per traceroute for each of the countries we study.}
        \label{fig:unlabeled_cdf}
    \end{figure}
}

\newcommand{\urlstable}{
    \begin{table*}[t]
    \centering
    \begin{tabular}{|l|r|r|r||r|r|r|r|}
\hline
\multicolumn{1}{|c|}{\multirow{2}{*}{Country}} &
  \multicolumn{3}{c||}{URLs} &
  \multicolumn{3}{c|}{Probed} &
  \multicolumn{1}{c|}{\multirow{2}{*}{Traceroutes}} \\ \cline{2-7}
\multicolumn{1}{|c|}{} &
  \multicolumn{1}{c|}{Official} &
  \multicolumn{1}{c|}{Certificates} &
  \multicolumn{1}{c||}{Validated} &
  \multicolumn{1}{c|}{Domains} &
  \multicolumn{1}{c|}{IPs} &
  \multicolumn{1}{c|}{Vantage Points} &
  \multicolumn{1}{c|}{} \\ \hline
    Belarus & 100 & 253 & 91 & 91 & 83 & 9 & 819 \\ \hline
    Brazil & 524 & 614 & 102 & 100 & 123 & 10 & 989 \\ \hline
    Canada & 121 & 630 & 100 & 100 & 115 & 10 & 998 \\ \hline
    China & 40 & 171 & 101 & 99 & 253 & 9 & 885 \\ \hline
    Israel & 229 & 225 & 101 & 99 & 105 & 8 & 792 \\ \hline
    Kenya & 83 & 135 & 100 & 100 & 88 & 9 & 849 \\ \hline
    Philippines & 62 & 116 & 109 & 99 & 144 & 10 & 984 \\ \hline
    Singapore & 120 & 225 & 196 & 100 & 115 & 10 & 994 \\ \hline
    Taiwan & 168 & 280 & 101 & 99 & 223 & 10 & 988 \\ \hline
    USA & 18762 & N/A & 100 & 100 & 119 & 10 & 998 \\ \hline
    UAE & 45 & 120 & 98 & 98 & 120 & 10 & 976 \\ \hline
    \end{tabular}
    \caption{For each country, we curate a list of potential target domains from official sources, expand the set using certificates, and manually validate each. We select up to 100 domains to probe from up to 10 vantage points resulting in almost 1,000 traceroutes before data sanitization.}
    \label{tab:urls-per-country}
    \end{table*}
}

\newcommand{\tracerouteresultspercountry}{
    \begin{table*}[t]
    \centering
    \SetTblrInner{rowsep=2pt,colsep=3pt}
    \begin{tblr}{|l|r|r|r|r|r|r|r|}
    \hline
    \SetCell[r=2]{c} Country & \SetCell[r=2]{c} {Total \\ Traceroutes} & \SetCell[r=2]{c} {Non-Anycast \\ Traceroutes} & \SetCell[r=2]{c} {Convergent\\(\%)} & \SetCell[r=2]{c} {Divergent\\(\%)} & \SetCell[c=3]{c} \% CitM of & & \\ \hline
    & & & & & Overall & Convergent & Divergent \\ \hline
    Belarus & 801 & 801 & 98.88 & 1.12 & 0.00 & 0.00 & 0.00 \\ \hline
    Brazil & 896 & 824 & 96.36 & 3.64 & 8.13 & 7.30 & 30.00 \\ \hline
    Canada & 899 & 791 & 91.78 & 8.22 & 22.63 & 21.49 & 35.38 \\ \hline
    China & 687 & 687 & 98.69 & 1.31 & 0.58 & 0.59 & 0.00 \\ \hline
    Israel & 792 & 606 & 73.60 & 26.40 & 14.69 & 0.45 & 54.38 \\ \hline
    Kenya & 849 & 791 & 63.66 & 36.34 & 37.97 & 10.83 & 85.52 \\ \hline
    Philippines & 984 & 636 & 21.07 & 78.93 & 43.08 & 34.33 & 45.42 \\ \hline
    Singapore & 994 & 735 & 84.22 & 15.78 & 15.24 & 14.22 & 20.69 \\ \hline
    Taiwan & 988 & 938 & 94.46 & 5.54 & 15.57 & 11.96 & 76.92 \\ \hline
    USA & 998 & 719 & 100.00 & 0.00 & 0.14 & 0.14 & 0.00 \\ \hline
    UAE & 976 & 672 & 91.96 & 8.04 & 33.18 & 31.23 & 55.56 \\ \hline
    \end{tblr}
    \vskip 1em
    \caption{Percent of traceroutes to convergent 
    and divergent 
    destinations \textit{after data sanitization}, and percent of traceroutes with \violators. Percentages are relative to the number of non-anycast traceroutes.}
    \label{tab:results-per-country}
    \end{table*}
}

\newcommand{\reasonsforbad}{
    \begin{table*}[t]
    \centering
    \SetTblrInner{rowsep=2pt,colsep=2.5pt}
    \begin{tblr}{
        colspec = {|l|r|r|r|r|r|r|r|},
        cell{3}{4} = {yellow!50}, 
        cell{4}{3} = {yellow!50}, 
        cell{4}{6} = {yellow!50}, 
        cell{4}{7} = {yellow!50}, 
        cell{5}{5} = {yellow!50}, 
        cell{6}{5} = {yellow!50}, 
        cell{6}{7} = {yellow!50}, 
        cell{7}{6} = {yellow!50}, 
        cell{9}{6} = {yellow!50}, 
        cell{12}{6} = {yellow!50}, 
    }
    \hline
    \SetCell[r=2]{c} Country & \SetCell[r=2]{c} {Total Traceroutes} & \SetCell[c=3]{c} {Traceroutes Discarded} & & & \SetCell[c=2]{c} {Traceroutes Modified} & & \SetCell[r=2]{c} {Subtotal} \\ \hline
    & & No Info & Website & Probe &  Geolocation & \SetCell{c} Squatting & \\ \hline
    Belarus & 819 & 0 & 18 & 0 & 0 & 0 & 18 \\ \hline
    Brazil & 989 & 93 & 0 & 0  & 1 & 18 & 112 \\ \hline
    Canada & 998 & 0 & 0 & 99 & 0  & 0 & 99 \\ \hline
    China & 885 & 0 & 0 & 198 & 0  & 23 & 221 \\ \hline
    Israel & 792 & 0 & 0 & 0 & 5 & 0 & 5 \\ \hline
    Kenya & 849 & 0 & 0 & 0 & 0 & 0 & 0 \\ \hline
    Philippines & 984 & 0 & 0 & 0 & 40 & 0 & 40 \\ \hline
    Singapore & 994 & 0 & 0 & 0 & 0 & 0 & 0 \\ \hline
    Taiwan & 988 & 0 & 0 & 0 & 0 & 0 & 0 \\ \hline
    USA & 998 & 0 & 0 & 0 & 9 & 0 & 9 \\ \hline
    UAE & 976 & 0 & 0 & 0 & 0 & 0 & 0 \\ \hline
    \end{tblr}
    \vskip 1em
    \caption{The number of traceroutes discarded or modified per country. 
    \textnormal{\textit{No Info} indicates that an empty traceroute is returned; \textit{Website} indicates the website is not an actual government website; \textit{Probe} indicates we do not trust the self-reported geolocation for the probe;  \textit{Geolocation} indicates that some hops seem to be improperly geolocated; and \textit{Squatting} breaks out traceroutes containing hops whose geolocation errors we suspect are due to the use of IP addresses commonly used for IP squatting. Cells with non-zero values are highlighted.}}
    \label{tab:reasons-for-bad}
    \end{table*}
}

\newcommand{\relatedworktable}{
    \begin{table*}[t!]
    \small
    \SetTblrInner{rowsep=2pt,colsep=1.35pt}
    \begin{tblr}{
        colspec = {|Q[l,m]|X[c,m]|X[c,m]|X[c,m]|Q[c,m]|X[c,m]|},
        width = \textwidth,
        cell{2}{2} = {red!20},
        cell{4-7}{2-3} = {red!20},
        cell{2}{4-6} = {red!20},
        cell{5-7}{4} = {red!20},
        cell{3-6}{5} = {red!20},
        cell{4-7}{6} = {red!20},
        cell{8}{2-6} = {YellowGreen!30},
        cell{3}{2-4} = {YellowGreen!30},
        cell{2}{3} = {YellowGreen!30},
        cell{4}{4} = {YellowGreen!30},
        cell{7}{5} = {YellowGreen!30},
    }
    \hline
    \SetCell{c} Paper & {Validates Hop \\ Geolocation} & {Validates VP \\ Geolocation} & {Considers Unreliable Responses} & {Considers Unreachable \\ Traceroutes} & {Considers Anycast Sites} \\ \hline
    Gupta 2014~\cite{gupta2014peering} & \cellcolor[HTML]{F4CCCC}No & \cellcolor[HTML]{D9EAD3}Yes & \cellcolor[HTML]{F4CCCC}No & \cellcolor[HTML]{F4CCCC}No & \cellcolor[HTML]{F4CCCC}No \\ \hline
    Fanou 2015~\cite{fanou2015diversity} & \cellcolor[HTML]{D9EAD3}Yes & \cellcolor[HTML]{D9EAD3}Yes & \cellcolor[HTML]{D9EAD3}Yes & \cellcolor[HTML]{F4CCCC}No & N/A \\ \hline
    Shah 2016~\cite{shah2016towards} & \cellcolor[HTML]{F4CCCC}No & \cellcolor[HTML]{F4CCCC}No & \cellcolor[HTML]{D9EAD3}Yes & \cellcolor[HTML]{F4CCCC}No & \cellcolor[HTML]{D9EAD3}No \\ \hline
    Edmundson 2018~\cite{edmundson2018nation} & \cellcolor[HTML]{F4CCCC}No & \cellcolor[HTML]{F4CCCC}No & \cellcolor[HTML]{F4CCCC}No & \cellcolor[HTML]{F4CCCC}No & \cellcolor[HTML]{F4CCCC}No \\ \hline
    Gueye 2018~\cite{gueye2018prevalence} & \cellcolor[HTML]{F4CCCC}No & \cellcolor[HTML]{F4CCCC}No & \cellcolor[HTML]{F4CCCC}No & \cellcolor[HTML]{F4CCCC}No & \cellcolor[HTML]{F4CCCC}No \\ \hline
    Candela 2021~\cite{candela2021worldwide} & \cellcolor[HTML]{F4CCCC}No & \cellcolor[HTML]{F4CCCC}No & \cellcolor[HTML]{F4CCCC}No & \cellcolor[HTML]{D9EAD3}Yes & \cellcolor[HTML]{F4CCCC}No \\ \hline
    Current Work & \cellcolor[HTML]{D9EAD3}Yes & \cellcolor[HTML]{D9EAD3}Yes & \cellcolor[HTML]{D9EAD3}Yes & \cellcolor[HTML]{D9EAD3}Yes & \cellcolor[HTML]{D9EAD3}Yes \\ \hline
    \end{tblr}
    \vskip 1em
    \caption{Comparison of how prior work addresses the challenges we identify in our paper (excluding validating government websites, as that is not applicable for these works).}
    \label{tab:related-work}
    \end{table*}
}

\begin{document}

\title{Country-in-the-Middle: Measuring Paths between People and their Governments}

\author{Alisha Ukani}
\affiliation{
    \institution{UC San Diego}
}

\author{Katherine Izhikevich}
\affiliation{
    \institution{UC San Diego}
}

\author{Shambhavi Mittal}
\affiliation{
    \institution{UC San Diego}
}

\author{Manan Patel}
\affiliation{
    \institution{UC San Diego}
}

\author{Samvrit Srinath}
\affiliation{
    \institution{UC San Diego}
}

\author{Kristy Ly}
\affiliation{
    \institution{UC San Diego}
}

\author{kc claffy}
\affiliation{
    \institution{CAIDA, UC San Diego}
}

\author{Alex C. Snoeren}
\affiliation{
    \institution{UC San Diego}
}

\raggedbottom

\begin{abstract}
Understanding where Internet services are hosted, and how users reach them, has captured the interest of government regulators and others concerned with the privacy of data flows.  In this paper we focus on government websites---services which arguably merit a higher expectation of protection against foreign surveillance or interference---and seek to identify countries in the middle (CitMs): countries that are neither the source nor destination in a path for a resident visiting their online government services. Finding these CitMs raises daunting methodological challenges.
We propose a framework to identify CitMs and use a pilot study of 149 countries to refine our methodology before conducting an in-depth measurement study of 11 countries.  For our focused study, we compile an extensive set of websites hosting government services and analyze over 9,000 IP-level paths from vantage points in those countries to these services. We conduct extensive manual validation to corroborate or discard paths based on the aforementioned challenges, and discuss paths that experience unexpected CitMs.


\end{abstract}

\maketitle

\section{Introduction}\label{sec:intro}

Data sovereignty has become increasingly important as countries fear their network traffic may be surveilled---or even tampered with---by foreign nations. 
%
Internet traffic that enters a foreign country becomes subject to that country's laws, which may stipulate that the government can gain access under certain conditions.  For example, 
Belarus requires domestic telecom providers to allow various levels of governmental access, including the ability to surveil (potentially foreign) traffic that transits their routers~\cite{belarus-amnesty}.
Such concerns are not hypothetical; in 2022, researchers found that a Russian state-controlled telecom company was routing traffic from Ukrainian networks through Russia~\cite{satariano22russia}, and, in 2013, Snowden revealed that GCHQ, a UK intelligence agency, was intercepting traffic on more than 200 fiber-optic cables~\cite{macaskill13gchq}.

Traffic may enter a foreign nation for a variety of political and technical reasons, many of which are benign---for example, due to submarine cables that were built between colonial powers and their former colonies~\cite{underseanetwork}; geographic proximity to other countries; and the use of cloud providers, large transit providers, and Internet exchange points (IXPs).
A particular foreign nation that receives traffic may be perceived as positive or negative depending on a country's political relationships.
In this work we do not take a stance on the desirability (or lack thereof) of any nations.
Regardless of the reason why traffic transits a foreign nation, if a government wants to manage the sovereignty of their data, they must first understand where their traffic is going (even if they ultimately do not change the locality of this traffic).
In particular, 
we focus on one class of Internet traffic where we expect that governments have greater visibility and would like to assess any potential security concerns:
traffic between a given country's residents and governmental websites.

Each nation must determine which governmental services to host domestically and which to host abroad. Presumably, such decisions are taken with full knowledge of the identity and applicable regulations of the nation selected to host the service (although the increasing prevalence of IP anycast raises doubts on even that point). It is less clear, however, whether a nation's decision makers are aware of how traffic will be routed between their residents and selected hosting locations.
If a government website is hosted within the country itself (which we term a \emph{convergent} destination), it is counterintuitive that a resident's traffic to that website would leave the country. Moreover, even if a website is hosted in a foreign country (which we call a \emph{divergent} destination), it is frequently unclear---and even time-varying---which other countries a resident's traffic may transit.

We consider how to answer the question: when residents located in a given country access their government's websites, what other countries are involved, with what frequency, and why? We refer to any country that is neither the source nor destination country of an IP-level path as a \emph{country in the middle (\violator)}. \violators can appear in paths to both convergent and divergent destinations.  (Prior work sometimes refers to paths to convergent destinations with \violators as ``boomerang'' or ``tromboning''~\cite{edmundson2018nation,fanou2015diversity,gueye2018prevalence,obar2012internet}.)  Figure~\ref{fig:definitions} depicts how we taxonomize the paths we study.
%


\ifshort
\else
\definitionsfigure
\fi

Despite the alluring simplicity of our question, our repeated attempts over multiple years to answer it have identified five core challenges that must be addressed in order to systematically identify \violators. The first two challenges are compiling a comprehensive set of government websites and finding sufficient in-country vantage points.
These challenges are manageable for studies of individual countries (e.g., one could deploy new vantage points, or select only countries with authoritative sources for government websites), but become increasingly daunting as the number of countries increases. 
Next, any study that relies upon in-band measurement of IP-level paths (which is the only practical approach at scale) must address the reality that not all IP hops are responsive to traceroute or similar tools, raising nuanced questions regarding how to interpret incomplete path information.
Finally, identifying \violators presents two additional challenges: accurately geolocating core infrastructure and identifying anycast destinations, for which it may be difficult to determine the replica in use. 

In this paper, we describe how these challenges frustrated our attempts to conduct a global-scale study that collects traceroutes from RIPE Atlas probes located in almost 150 countries around the globe.  We use concrete examples from our experience to motivate the design of a rigorous validation methodology that allows us to draw conclusions regarding \violator prevalence in a more focused study.
Concretely,
we develop a framework to study \violators and conduct a study of 11 countries in which we use 10 RIPE Atlas probes located in each country to launch traceroutes towards approximately 100 government websites for each country, resulting in a dataset of 9,278 successful\footnote{We fall short of the desired 11,000 traceroutes because some Atlas probes malfunctioned and some collected traceroutes our validation process determines to be unreliable.} traceroutes. We geolocate the intermediate hops to the best of our ability to identify \violators. Because of the limitations of available geolocation approaches, we cross-validate the geolocation of each alleged \violator hop.
While we believe our conclusions are sound, navigating the challenges requires extensive manual effort. 

Many of our validated \violators are plausible yet unexpected, either due to the geography of the countries involved, or because our dataset contains evidence of alternative paths between the same vantage point and the website without \violators. 
Operators may have good reasons for routing traffic through other ASes or IXPs, such as to avoid congestion---even if this may introduce \violators---but
we expect that some governments may not be aware of the set of nations through which their traffic may be routed.

The contributions of our work include:
\begin{itemize}
    \item Detailing the myriad challenges facing a systematic study of governmental traffic sovereignty, developed through a pilot study of \violators for government websites of almost 150 countries;
    \item Proposing a framework for studying \violators;
    \item Compiling lists of government websites for 11 countries and measuring IP-level paths to these services from in-country vantage points, resulting in a dataset of over 9,000 traceroutes;
    \item Performing extensive manual validation during data collection and post-processing;
    \item Extracting insights about common \violators for the 11 studied countries; and
    \item Recommending approaches to improve the coverage and reliability of future studies
\end{itemize}

\section{Challenges}\label{sec:challenges}

In this section, we enumerate the five significant challenges to conducting a study that seeks to uncover threats to traffic sovereignty.


\subsection{Compiling government websites}
There is no single authoritative source that provides a current, comprehensive list of official government websites for most countries. While the United Nations publishes a list~\cite{unportal}, it contains between one and four URLs per country, far too few to constitute a representative sample.
Some individual countries publish official lists, but we find many to be outdated and/or inaccurate.  In particular, websites on these lists may not actually host content, or---if the list is outdated---may no longer belong to the government. 

\subsection{Finding vantage points}
To measure IP-layer paths, we need vantage points distributed throughout the countries of interest that can run traceroutes to target websites. RIPE Atlas provides many vantage points in a large set of countries (though some countries have few probes) at low cost, but some of these vantage points are deployed in data centers and thus may not be representative of the Internet providers used by residents of a given country. Recent analysis has also found that some RIPE Atlas probes may misreport their geolocation~\cite{izhikevich2024trust}, forcing us to scrutinize their purported locations.

\subsection{Interpreting traceroute responses}\label{sec:challenges:traceroute}
The inferential challenges in interpreting traceroute data are
well documented \cite{luckie2016bdrmap}.
Routers may filter ICMP traffic, not respond (resulting in unlabeled hops
in traceroute outputs), or may respond with bogon (i.e., reserved or unallocated) IP addresses. A more subtle problem is that a router may respond using an IP address of an interface different from that which received the packet, or even an IP address that belongs to the interconnecting router for that interface.  
However, a 2014 study~\cite{2014-luckie-slda} found such responses to be relatively rare, and that traceroute remains the best (only) available tool for path analysis.

Similarly, tunneling techniques like Multiprotocol Label Switching (MPLS) may complicate interpreting traceroute responses by hiding hops inside a tunnel, which in turn may obscure potential \violators. A 2012 study~\cite{donnet2012revealing} found that such opaque tunnels that hide hops were rare in the wide-area Internet, and that it was far more common for MPLS tunnels to reveal the routers traversed within the tunnel.  That said, cloud providers use such techniques extensively~\cite{andromeda}, so future work could assess the extent to which our methodology under-reports \violators due to tunneling in private backbones (Section~\ref{sec:conclusion}).



\subsection{Performing IP geolocation}\label{sec:challenge:geo}
A fundamental task when attempting to map network paths to countries is geolocating individual IP addresses. Publicly available commercial geolocation databases and research-oriented geolocation services~\cite{maxmind,ipmap,ipinfo,hoihopaper,netacuity} vary tremendously in accuracy, precision, stability, and coverage~\cite{replicationgeo, ipmap,  Geo-database-comparison, gouel2021ip}, and are generally not focused on core Internet routers. As a result, these databases must be treated with skepticism and corroborated using other approaches.

\subsection{Dealing with anycast}
Anycast hosting allows multiple physical servers to present the same public IP address, which poses a challenge to geolocation services that map IP addresses to a single geographic area.
Because anycast can obscure the identity of the country hosting the website, it is difficult to determine if a government website is convergent or divergent---or both.  Moreover, in the case of divergent destinations, it complicates the validation and interpretation of \violators because the geographical relationship between the source and destination countries may be unclear.  
In our work we use an anycast-specific geolocation dataset~\cite{anycast_census} to identify potential countries hosting the anycast IPs we uncover, but the dataset is unfortunately not contemporaneous with our measurements (see Section~\ref{app:anycast} for details), so we do not incorporate the results into our main findings.


\section{Baseline Methodology}\label{sec:method}

To begin, we present a(n admittedly incomplete) methodology that navigates as many of these challenges as possible in an automated fashion before or during data collection.  We employ this methodology to conduct a global-scale pilot study and use the results to design an enhanced methodology that incorporates manual validation steps where necessary.

\methodology

\subsection{Collecting Government Websites}\label{sec:method:domains}


Given the lack of a single authoritative source, we compile lists of government websites on a per-country basis using a three-step process depicted in Figure~\ref{fig:methodology}: (1) We start by combining the few government websites provided by the United Nations~\cite{un_nsgt} with additional sites we obtain from lists of government websites for each country.  From each of these URLs, we extract the relevant domain, i.e., eTLD+1, to use as a seed.  
(2) We identify candidate secure websites within these official domains by consulting the Censys TLS certificate repository~\cite{censys15} and extracting URLs contained within the certificates for seed domains. (3) We arrive at a final set of target domains
by confirming the liveness of each candidate website.


\subsubsection{Official Government Websites}\label{sec:method:officialgov}

To expand upon the few sites listed in the UN website, we compile lists of official government websites for each country collected from various sources. The U.S. government publishes expansive datasets with this information~\cite{dot_gov_domains,military_domains}.
Because manually collecting similar lists for all nations is
prohibitively labor intensive, we instead devise an automated approach.  After
manually correcting some inaccuracies in the U.N. list and adding some well-known
domains for other nations we
%
use them to bootstrap a search for additional governmental websites. 

\subsubsection{Expanding our Set of Domains}\label{sec:method:censys}

Prior work has relied on DNS logs~\cite{houser2022comprehensive} or fixed sets of localized hostname patterns like \texttt{*.gov.ccTLD} (and other variations in different languages) to find government websites~\cite{rkumar:imc24,singanamalla2020accept}.
Instead, we employ Censys~\cite{censys15} to search for subdomains of our seed domains (and their eTLD+1's) in active TLS certificates.
Our approach of extracting eTLD+1's from authoritative domains allows us to not only infer the patterns used in prior work, but also discover eTLD+1's that do not match those patterns.  For example, we find Canada's primary government eTLD+1 (\texttt{gc.ca}) does not match the patterns used in prior work~\cite{singanamalla2020accept}. 
Finally, the presence of a TLS certificate suggests that the website operator is following best security practices and that they would like to avoid security concerns like machine-in-the-middle attacks.


\subsubsection{Verifying Liveness}

The existence of a certificate, however, does not indicate that there
is a corresponding operational web service. 
We confirm the existence of a website by resolving the fully qualified domain name (FQDN) and attempting to contact a Web server at that address.

\subsection{Measurement Vantage Points}\label{sec:method:vantage}

We conduct our measurements from RIPE Atlas probes, a large, distributed set of vantage points for measurement research.
We select up to 10 RIPE Atlas probes
per country that maximize AS diversity; if a country has fewer than 10
RIPE Atlas probes, we use them all. 
Unfortunately, many countries do not have a robust
infrastructure of RIPE Atlas probes, reducing our flexibility.
We also include data center and anchor probes (which are typically hosted by companies as opposed to individuals); future work should consider whether hardware probes more accurately reflect residential Internet service provider paths.

\subsection{Path Collection}\label{sec:method:traceroute}

We run ICMP (Paris) traceroutes from the selected probes in each country to the set of target domains for that country.  (We discuss the implications of using ICMP vs TCP in Section~\ref{sec:highlevel:reachability}.)
%
We perform DNS resolution on the probe itself to mimic user traffic; this frequently results in different vantage points probing different IP addresses for the same domain.
%
We use IPinfo~\cite{ipinfo}, a geolocation database with higher accuracy compared to others~\cite{replicationgeo}, to geolocate each hop including the destination.



\section{Preliminary Study}

We conducted a pilot study in May 2023 that attempted to be as expansive as possible.
We refine our methodology from the pilot study for the subsequent focused study (Section~\ref{sec:focused});
we share this pilot study in the interest of guiding future researchers through the many challenges of conducting a large-scale study of \violators.

\subsection{Methodological details}

We followed the baseline methodology from the previous section with a few refinements (that we subsequently revise in our focused study as detailed in Section~\ref{ss:md}).

\subsubsection{Website filtering}

Starting from the United Nations list of websites and a few other seed domains, our Censys search yielded 763,543 FQDNs
across 110,068 registered domains, 388 eTLDs, and 196 countries.  
We filtered the list by using ZDNS~\cite{zdns} with the 8.8.8.8 name
server to focus only on domains that successfully resolve.  We then used ZMap~\cite{zmap}
to ensure port 443 was open on at least one IP address per FQDN, increasing our confidence that there is an operating Web service---but
we did not inspect the content available.
This filtering process resulted in 497,475 FQDNs spanning 93,548 registered domains and 362 eTLDs.

Resource constraints necessitated sampling among the available FQDNs to select our measurement targets; 
we sought to maximize the diversity of destination IP
addresses.
We used the CAIDA
prefix2as dataset~\cite{pfx2as} to find the corresponding IP prefix and AS
announcing the IP address associated with each FQDN in our list.  We chose up to 100 FQDNs per country that maximize,
in order, 1) number of ASes, 2) IP prefix diversity, and 3) number of unique IP
addresses.
This yielded 12,450 FQDNs for 194 countries.
%
While we attempted to collect 1,000 measurements per country (10 probes each
visiting 100 domains) we fell (far) short: our
measurement campaign yielded 71,164 traceroutes for 149 countries.

\subsubsection{Post-processing}

We employ a variety of filtering techniques to
increase our confidence in the geolocation data we consider.
Within a traceroute, we
first strip out all hops in private IP spaces, as it is impossible to
get any geolocation information for these addresses (6,633 hops across
64.12\% of traceroutes). To reduce noise, we filter hops where
the IP address geolocates to a country code that only appears once in
the entire traceroute, corresponding to 2,779 hops across 17.50\% of
traceroutes (we revise this strategy in the subsequent focused study).
We then remove any traceroutes where, as a result of the previous filters, we are
left with no hops between the source and destination IP addresses; this is the
case for 1,042 traceroutes. 
After this filtering process, we are left with 68,764 traceroutes for
149 countries.

\BanjoGraphOld

\subsection{Initial results}

Figure~\ref{fig:banjo_old} presents the results of our pilot study, where each country is represented by a stacked horizontal bar chart that plots paths to convergent targets to the left of zero on the $x$ axis, while divergent targets are plotted to the right. Traceroutes to anycast targets (identified using the MAnycast$^2$ dataset~\cite{sommese2020manycast2} collected on January 2022) are graphed separately on the far right. The bars are color-coded by the number of \violators in the path and normalized according to the number of probes used in that country, so the length of the bar corresponds to the average number of traceroutes collected per probe---which is expected to be 100, but is frequently less due to measurement errors and/or lack of target domains for that country.  Yet, one immediate takeaway from the pilot study is that even for countries where our Censys-based target selection is unable to generate many targets (i.e., those countries with very short bars), we still uncover \violators (e.g., Liechtenstein, Vanuatu, Montenegro, etc.).  


At a high level, these results are consistent with intuition: for example, in general there are more \violators on paths to divergent targets than convergent ones (i.e., the bars to the right of zero have more pink and orange than those to the left).  That said, we do indeed find \violators on paths to convergent targets, suggesting that
even if a government hosts their online services
domestically, there is no guarantee that traffic from residents to these
services will stay local. 
Moreover, many of these instances survive closer scrutiny.

\subsubsection{Convergent Traceroutes}

One explanation for the presence of \violators for convergent paths is that the Atlas probe is connected
to a foreign-owned network, which may route domestic traffic through other
countries. In our dataset, we find several instances of this happening. For
example, we find probes in Mexico and Russia connected to the
``ANEXIA Internetdienstleistungs GmbH'' AS organization, which then peers with
Arelion (a telecom based in Sweden), and so some traceroutes have Swedish
hops. 
Another example is that a Brazil probe connected to the U.S.-owned
``Neustar Security Services'' network then sends traffic to U.S. hops in the
U.S.-owned ``Tata Communication'' network. This case is particularly concerning,
as the destination website is the email log-in page for the Brazilian Air
Force, and Brazil has been vocal about combating U.S. surveillance of Internet
traffic~\cite{brazil}.

Another reason why convergent \violators may occur is that domestic ASes may
peer with foreign ASes. We again see many examples of this in our dataset. In
one case, the Iraqi network ``Zana Mohammed Mahdi A.Rahman company for Internet
Service Provider LTD'' peers with the U.S.-based network ``Cogent
Communications.'' In another example, we find a RIPE Atlas probe in Georgia,
connected to a Georgian network\footnote{The probe is connected to the Aquafon
telecom, which is in Abkhazia, a partially-recognized state that most countries
recognize as part of Georgia.} that routes traffic through Russia while heading
to the Georgia government website for visas. Specifically, the probe's
network ``ZAO Aquafon-GSM'' peers with the Russian AS ``Dmitriy V. Kozmenko.''




\subsubsection{Divergent Traceroutes}




Similarly, spot-checking a variety of divergent cases yield a number of frequent, plausible scenarios.
Some are geographical, where adjacent nations appear on routes to foreign targets (e.g., South Africa's appearance as a \violator on paths originating in Tanzania) or island nations that make use of submarine cables landing at multiple \violators on their way to the mainland.  Others seem to be artifacts of colonial relationships, which have historically affected the development of submarine cables~\cite{underseanetwork}. For example, Botswana has the U.K. as a \violator for all of its 16 divergent traceroutes. Similarly, Lebanon and Tunisia have France as a \violator for over 90\% of their divergent paths with \violators.

\subsection{False positives}

\urlstable

On the other hand, the results of the pilot study contain a number of \violators of which we are suspicious, based on geolocation information extracted from hostnames via the CAIDA Hoiho tool~\cite{hoihopaper}. These examples include:

\begin{itemize}
    \item The Czech Republic appears as a \violator in paths to nine different convergent websites for a single Japanese probe. These include sensitive websites such as a website to schedule COVID-19 vaccinations as well as subdomains of Japan's National Police Agency. However, upon further inspection, Hoiho geolocates all of these alleged \violator IP addresses to Japan, meaning these paths do not contain any actual \violators.
    \item Similarly, a single Canadian probe reports seven paths to convergent websites as having the U.S. as a \violator. Several of these websites are related to the Canadian government's VPN tool~\cite{cavpn}.
    While the U.S. is geographically close to Canada and thus plausible, Hoiho actually reports these IP addresses as residing in Canada. Again, this information eliminates all \violators for these paths.
    \item Finally, two paths to the login portal for the Reserve Bank of Australia appear to have the U.S. as a \violator, but according to Hoiho, one of the IP addresses allegedly in the U.S. actually geolocates to Australia.
\end{itemize}

We additionally find one inaccuracy in the U.S. list of non-\texttt{.gov} domains: the website \texttt{nc-ddc.org} was added in 2013 as the website for the North Carolina Council on Developmental Disabilities. However, when visited in 2023, this is a Japanese website with information on what is safe and unsafe to feed dogs. We submitted a pull request to remove this domain from the list of official non-\texttt{.gov} domains, but it suggests that even government websites from official datasets may not be trustworthy and must be manually inspected.

\section{Focused Study}\label{sec:focused}

The anecdotes above are but a few of the dubious \violators we discovered in our pilot study, causing us to refine our methodology to address a number of shortcomings that gave rise to many of the spurious datapoints.  Some aspects of validation (validating government websites, geolocations, and traceroute responses) require manual analysis and corroboration with external sources in post processing. 
(Appendix~\ref{app:valdetail} discusses how this post-processing impacts the conclusions we draw from our data set.)

\subsection{Country selection}

Due to the extensive manual validation we perform to confirm the data, we limit the scope of our reported results to 11 countries, each of which host at least 20 active RIPE Atlas probes.
Our selection criteria for countries is to maximize geographic diversity (including at least one country per continent) and diversity of  Internet Freedom score~\cite{freedomhouse}, a measure of a country's digital rights (e.g., freedom from censorship, right to Internet access). For the latter criteria, the selected countries have Internet Freedom scores roughly following an even distribution with a slight skew towards countries with greater freedom.
%

\subsection{Methodological refinements}
\label{ss:md}

To increase the number of seeds for our Censys search beyond the few sites listed in the UN website, we compile lists of official government websites for each country collected from various sources. We continue to use the official datasets for the United States~\cite{dot_gov_domains,military_domains} as well as for Brazil~\cite{brazil_domains}. For eight other countries, we find official, but much more limited, lists of websites for government organizations. For the remaining country we study, China, we were unable to identify any official source, so we use Wikipedia to find links to government agencies~\cite{china_wiki}.
We detail the specific sources used for each country in Appendix~\ref{appendix:gov_website_collection}. 

\subsubsection{Website filtering}

Unlike our pilot study, which only confirmed the existence of a webserver at target FQDNs, for our full study we deploy a VPN-hosted client within each country and manually inspect the returned webpage to ensure that it hosts content that belongs to the respective government. 
Concretely, to identify a target set of domains to probe, we manually verify whether each candidate URL hosts a government website to the best of our ability.
Similar to prior work~\cite{rkumar:imc24},
we use ProtonVPN, a popular VPN with endpoints in almost all of the studied countries, to open a Web connection from an endpoint within the relevant country to the candidate URL.  
If there is no ProtonVPN endpoint in the selected country, we choose a ProtonVPN endpoint in a neighboring country to visit those websites (e.g. we use the Hong Kong VPN endpoint to visit Chinese websites).
We acknowledge that this approach may result in discarding valid government websites that have strict geofencing or that block connections from VPNs.

If we are successful in contacting a webserver, we inspect the returned webpage (sometimes employing machine translation) to check if it belongs to 
an official government agency/body (both at federal and state/province levels, e.g. president and ministry websites), or a state-owned enterprise.
We repeat this process until we obtain at least 100 \emph{validated URLs} for each country, or exhaust our list of candidate URLs. We first examine URLs from the official sources and then randomly select among those obtained from our Censys search. 


\subsubsection{Post-processing}
\label{valid:geo}

In the pilot study, we conservatively filter out countries that only appear once in a given traceroute in an effort to reduce spurious \violators, but the heuristic is obviously imperfect.  As an alternative, we considered an automated verification method that compared the latency between the purported \violator hops and Atlas probes located in both the source country and the \violator.  Specifically, we collected \texttt{ping} measurements from up to five probes in both the source country and the alleged \violator and attempted to use Student t-tests to check for statistical differences in the ping latencies. Unfortunately, we find that the distribution of latencies from even a single probe may vary, and as previously mentioned, we cannot trust the self-reported geolocation of probes, mooting this approach.  Another technique we considered was to cross-validate IPinfo's geolocation with that reported by the CAIDA Hoiho tool~\cite{luckie2021learning}.  Unfortunately, Hoiho works based on hostnames, which are only available for
26\% of the IP addresses in our dataset.

Instead, in our focused study, we validate the geolocation of each IP address in our traceroutes that corresponds to a potential \violator using
a highly conservative RTT-based method from prior work~\cite{replicationgeo,izhikevich2024trust}. 
%
%
If the reported latency between a given hop and the source probe (as opposed to the prior hop~\cite{fontugne2017pinpointing}) is less than 2/3 the speed of light between the source and the hop's IPinfo-inferred country,
we consider IPinfo to be erroneous.
If the geolocation information for all of hops in a traceroute that were inferred to be located within a particular \violator are classified as errors using the aforementioned criteria, we \emph{modify} the traceroute by removing those hops (i.e., discard the \violator). We do not discard the traceroute entirely, nor do we attempt to correct the geolocation---likely resulting in an under-reporting of \violators for this type of traceroute. 

\reasonsforbad

\section{Data Sanitization}
\label{sec:navigating}

We collected the traceroutes used for our focused study in February 2024.  The number of URLs, vantage points, and traceroutes collected per country is reported in Table~\ref{tab:urls-per-country}.  We manually inspect each of the 2,088 IP hops corresponding to \violators in our dataset.
This section
details instances where we discard, or---when appropriate---modify traceroutes (by filtering out hops with invalid IP addresses).
Inspection of the data reveals that one Brazilian probe (7113) returned empty responses for every traceroute so we discard all 93 paths collected by the probe.
More fine-grained inspection discards or modifies an additional 411 traceroutes.  We summarize the quantities in Table~\ref{tab:reasons-for-bad} and detail the reasons below. 


\subsection{Verifying Government Websites}\label{sec:navigating:website}

Unfortunately, our manual process to validate government websites is not foolproof, especially when foreign languages are involved.
Our initial set of 100 validated domains for Belarus includes \texttt{dha.by} which we extracted from a list of official Belarusian websites published by the Belarus Ministry of Foreign Affairs~\cite{belarus_domains}. We discovered in subsequent analysis that this website does not, in fact, belong to the government. 
Hence, we discard paths to it in post-processing.  We similarly discard paths to a Belarusian social media site that was similarly listed among its governmental websites.  

\subsection{Filtering Vantage Points}\label{sec:navigating:probes}

A recent study reported that some RIPE Atlas probes misreport their own location~\cite{izhikevich2024trust}.  In particular, by comparing RTTs from the probe's self-reported geolocation to servers with known geolocations, the authors identify violations of the 2/3 speed-of-light threshold.  We find one Canadian probe (6493) in our dataset on their published list of improperly located Atlas probes and discard all traceroutes collected from this probe in our dataset.
We also
discover that two of the allegedly Chinese probes (7030 and 50179) we use are actually located in Hong Kong according to their self-reported latitude/longitude, which we analyze separately for the purposes of this study;
hence, we discard paths from these probes.

\subsection{IP squatting}

Almost half of the improperly geolocated hops we discover in our dataset are due to IP squatting, so we call them out separately in Table~\ref{tab:reasons-for-bad}.
IP squatting refers to the use of IPv4 addresses that were historically allocated but not announced.  
We see 41 traceroutes in Brazil and China that experience \violators for hops in this the 11.0.0.0/8 IP address range assigned to the U.S. DoD Network Information Center (AS749), which prior work found is often used for IP squatting~\cite{salamatian2023squats}.
Specifically, four China probes (all hosted within Alibaba) report traceroutes containing such DoD IP addresses---implying the U.S. as a \violator---for 25 traceroutes. 
We modify these 25 traceroutes by discarding the hops corresponding to IP addresses in this subnet. Similarly, 18 traceroutes from Brazilian vantage points also include hops in the 11.0.0.0/8 range; we again modify these traceroutes by discarding these hops. 
We may miss additional cases of IP squatting, especially for IP addresses announced by smaller organizations.

\subsection{Interpreting Traceroute Responses}\label{sec:navigating:traceroute}

\unlabeledcdf

We explore the impact of two limitations of traceroute: unreliable responses and paths that do not reach the destination.




\subsubsection{Unlabeled hops}

Unlabeled hops in a traceroute (i.e., IP hops that do not generate ICMP responses) may hide additional \violators. This means our study provides a lower bound on the number of \violators experienced by each country.  In order to get a sense of the scale of missing data, we plot a CDF of the longest set of consecutive unlabeled hops for each country in Figure~\ref{fig:unlabeled_cdf}. This figure presents two main takeaways. First, there is pervasive missing data: no more than 40\% of any country's traceroutes are complete. Future work has the opportunity to incorporate information before and after unlabeled hops (as well as other information like BGP announcements) to infer what these hops may be, but this is out of scope for our work. Second, countries are associated with different levels of transparency in the infrastructure they depend on to reach their government services. For example, the Philippines has the highest transparency in our dataset (61\% of traceroutes do not have any consecutive unlabeled hops, i.e. values of 0 or 1 for the $x$-axis of Figure~\ref{fig:unlabeled_cdf}), while Canada has the lowest transparency (86\% of traceroutes have consecutive unlabeled hops).

\tracerouteresultspercountry

\subsubsection{Traceroute reachability}
\label{sec:highlevel:reachability}

Another limitation of traceroute is that many network devices filter out ICMP traffic, so some of our traceroutes are unable to reach the destination website. Only 57\% of our traceroutes (5,888) reach the destination IP, with a very slight improvement to 58\% of traceroutes (5,935) reaching the destination AS (i.e., the traceroute stops recording hops within the AS that originates the destination IP address).
We do not exclude these results from our dataset because that could bias our data, e.g., we do not want to systematically exclude studying destinations hosted by Amazon (a hosting provider we find to frequently filter ICMP in our dataset), especially if we can still identify \violators before the traffic is filtered.

\ifshort
\else

That said, we acknowledge that the 3,929 traceroutes that do not reach the destination AS could under-report \violators.  As a way to characterize the portion of paths we miss, we attempt infer the missing ASes through which the traceroute could have traversed by comparing the set of ASes identified on the traceroute with AS paths in BGP advertisements.  Using the February 2024 routing-table snapshots from Routeviews~\cite{routeviews} and RIPE RIS~\cite{ris} projects, we first remove BGP routes that had been observed for less than five days to avoid spurious paths. We extract the BGP routes whose AS paths contain both the source and destination ASes from at least one of our incomplete traceroutes.  We are able to match 410 traceroutes (i.e., about 10\% of the incomplete traceroutes) to a corresponding BGP route, identifying 66 additional ASes that our traceroutes could have traversed.
The top-five ASes that did not respond to the most traceroutes are a cloud hosting provider (Amazon) followed by ASes run by governments (Canada) and Local ISPs in Singapore and Israel.  While interesting, this BGP analysis is only circumstantial, so we do not include these findings in the results reported in subsequent sections.
\fi

In an effort to determine whether our reachability results might be biased by the use of ICMP---as opposed to the transport protocols used by typical Web traffic---we ran a brief follow-up study in March 2025 to compare ICMP, TCP, and UDP traceroutes for two countries: Kenya and the U.S. For both countries, TCP traceroutes reach the destination IP address more frequently than ICMP (92\% vs 80\% for the U.S., and 87\% vs 75\% for Kenya)\footnote{In 2024, 78\% of paths from the U.S. and 70\% of paths from Kenya reached. Other countries had lower reachability rates, lowering the average for 2024.} or UDP (which only reached the destination 24\% of the time in the U.S. and 28\% for Kenya).
Manual investigation reveals that when ICMP is filtered and stops short of the final few hops revealed by TCP, these last hops are often within the destination AS and do not reveal any additional \violators.

Perhaps more interestingly,
there are a few cases where ICMP and TCP traceroutes for the same source/destination pair are routed through different transit providers.  Yet, even in those cases,  we find that both protocols traverse the same set of \violators (which are furthermore the same as those seen in our main study data collected in February 2024).  Hence, while we acknowledge that TCP traceroutes may provide greater coverage---and potentially additional \violators---we believe that the \violators revealed through our ICMP traceroutes are salient.
Future work could explore comprehensively comparing traceroutes from both protocols.

\section{Results}\label{sec:unrefined}

Table~\ref{tab:results-per-country} presents the fully sanitized (i.e., validated, post-processed, and manually inspected) results of our focused measurement study, showing the percentage of convergent and divergent traceroutes for each country as well as the prevalence of \violators.  Figure~\ref{fig:banjo} visualizes these results in the style of Figure~\ref{fig:banjo_old}, with the addition of outlines surrounding each stacked bar that show the total volume of traceroutes collected, including those that were discarded during post-processing and/or sanitization (i.e., appear in Table~\ref{tab:reasons-for-bad}).  In this section we describe the high-level takeaways we extract from our results, as well as the negligible impact anycast seems to have on our findings.  Additional anecdotes we discovered as part of our manual inspection are included in Appendix~\ref{app:int}.

\banjograph

\subsection{Insights Gained}

We extract three high-level trends that may explain the patterns of convergence/divergence and \violators in our dataset. We emphasize that these are trends we observe, but each trend has some counter-examples.
\begin{enumerate}
    \item Countries with well-developed Internet infrastructure are more likely to host their websites domestically, but this relationship is not perfect. In fact, countries can take measures to increase their levels of convergence.
    \item The level of convergence vs divergence influences the amount of \violators a country experiences, as divergent traceroutes may necessarily transit \violators.
    \item Countries with higher levels of Internet infrastructure experience fewer \violators.
\end{enumerate}



\subsubsection{Rate of convergence}

We find meaningful variation in domestic vs. foreign hosting for each country, and find that countries that host more websites locally are often---but not always---those with robust Internet infrastructure.
%
Seven of the 11 countries have strong domestic hosting infrastructure, as more than 90\% of traceroutes reach convergent targets. Many of these countries (e.g. the U.S., Taiwan, China, Canada, and the U.A.E.) all have robust Internet infrastructure. However, one of these countries, Belarus, is not known for having a robust Internet infrastructure. We believe that the explanation for almost 99\% of paths reaching convergent targets is a presidential decree that ordered all \texttt{.by} domains must be physically hosted within the country~\cite{belarus-net-freedom}.


Similarly, countries with less developed Internet infrastructure often have more divergent targets. Countries like the Philippines and Kenya are still developing their infrastructure, and both countries have a high fraction of traceroutes to divergent websites.
However, both Israel and Singapore have a surprisingly high fraction of traceroutes to divergent destinations given their robust infrastructure.
Both countries host many of their government websites in the U.S., and Israel also hosts websites in Europe.



\subsubsection{Relating \violators to Geographic Factors}

Intuitively, one may expect that paths to convergent destinations experience fewer \violators than paths to divergent destinations, which may transit other countries out of geographic necessity. 
%
We observe some correlation between divergence and \violators in our dataset. Israel, Singapore, and Canada all experience many European \violators en route to web servers hosted in Europe. Brazil experiences nine traceroutes with the U.S. as a \violator en route to Canada. 
On the other hand, we see a number of \violators on paths to convergent destinations---some of which may be explained by geographic proximity.
The Philippines has the highest rate of \violators for convergent targets at 43\%; the most common \violator for these paths is Singapore, followed by Hong Kong. We now present two examples (with validation details) of geographically-proximal \violators to convergent websites.

\heatmapviolators

\paragraph{Hong Kong as a \violator for Taiwan}\label{interesting:tw}

Two of our Taiwan probes are hosted by PCCW Global, but one consistently experiences Hong Kong as a \violator, and the other does not.\footnote{Probe 6181 sees \violators, and probe 1002754 does not. Probe 6818's IP address is announced by both PCCW Global (AS3491) and Gateway Communications (AS31713). Gateway was acquired by PCCW in 2012.}
We find that almost 75\% of the traceroutes for one probe visit Hong Kong on the second hop before returning to Taiwan on the third hop (via an IP address announced by Chunghwa Telecom).
Meanwhile, the other probe also visits an IP address announced by Chunghwa Telecom in the second or third hop in almost 90\% of its traceroutes, but never experiences a \violator.

For the probe experiencing the \violators, 73 of the 99 traceroutes visit Hong Kong on the second hop and return to Taiwan on the third; on average, these traceroutes jump from a latency of 1.11~ms on the first hop to 47.25~ms on the second hop and 106.99~ms on the third hop. For its 19 traceroutes without any \violators, on average the first three hops have latencies of 1.12~ms, 6.35~ms, and 69.76~ms---each far lower than the previous set of latencies, which supports the geolocation inferences for Hong Kong.

In the case of the probe that does not experience Hong Kong as a \violator, 86 of its 99 traceroutes visit Chunghwa Telecom (the same AS that the first probe reaches on the third hop) on the second or third hop. On average, RTTs of the second and third hops are 0.77~ms and 1.62~ms. Again, these are far lower than the latencies experienced in the other probe's paths, which suggests that Hong Kong is a plausible \violator and not a geolocation error.




\paragraph{Hong Kong as a \violator for Singapore}\label{interesting:sg}

Of Singapore's 88 traceroutes to convergent websites with \violators, almost half (38) see Hong Kong as a \violator. We find that 15 of these paths transit the Hong Kong Internet Exchange (HKIX), all from the same probe that likely relies on HKIX for connectivity. We find another probe also consistently sees Hong Kong as a \violator for its fourth hop and
traceroute latencies spike from (on average) 1.96~ms on the third hop to 29.09~ms on the fourth hop, indicating the \violator is plausible.




\subsubsection{\violators vs. Internet Infrastructure}

Intuitively, countries with well-developed economies and Internet infrastructure are more likely to be \violators for many other countries.
Figure~\ref{fig:heatmap_violators} presents a heatmap to illustrate the frequency of \Violators to both convergent and divergent (i.e., not anycast) destinations.
France, Germany, Hong Kong, Singapore, and the U.S. are common \violators across countries around the world. Hong Kong is a frequent \violator for countries in Southeast Asia like the Philippines, Taiwan, and Singapore (as described above).

The heatmap also reveals strong relationships between certain pairs of countries. For example, Singapore is a common \violator for Indonesia, and the U.S. is a common \violator for Canada, both of which may be expected due to geographic proximity. However, it also reveals some unexpected pairings. For example, we see France and the U.K. as common \violators for Kenya despite being geographically distant. We also see Singapore and France as common \violators for the United Arab Emirates. We describe two case studies of \violators for Kenya and the U.A.E. in this section. While Singapore has many \violators in Africa and Europe, some of these are due to one probe (7219) hosted by Angola Cables, which sends all but one of its paths through South Africa; these \violators likely do not generalize to residents as Angola Cables does not advertise services in Singapore.

\paragraph{U.K. and South Africa as \violators for Kenya}\label{sec:interesting:kenya}

A decade ago, prior work found that Liquid Telecom routed traffic from South Africa to Kenya through London~\cite{gupta2014peering}. Interestingly, as Kenya has developed its infrastructure, its traffic that transits Liquid Telecom often avoids \violators. We observe two convergent destinations for which most probes (seven of eight) transit Liquid Telecom and do not experience \violators, but one probe does experience \violators while transiting Liquid Telecom.\footnote{Probe 13218 experiences these \violators for RIPE Atlas measurement IDs 68064916 and 68064948. One Kenyan probe malfunctions for these destinations, so only eight of the nine probes reported results.}
The latter probe goes through the U.K. and South Africa before returning to Kenya for both destinations.

We validated these geolocation results through Hoiho and latency analysis. Hoiho confirmed one of the four U.K. addresses, but did not have information about the three South African ones. Another U.K. IP address maps to Liquid Telecom's IP address in the London Internet Exchange (LINX), supporting this inference. The latency also jumps from Kenya to the U.K. in both traceroutes (0.54ms to 194.51ms in one traceroute and 0.81ms to 197.21ms in the other).

\paragraph{Singapore as a \violator for the U.A.E.}\label{interesting:uae}

Singapore is a \violator for 14\% of all paths to convergent destinations for the U.A.E. While Singapore is geographically far from the U.A.E., we cross-reference the 88 IP addresses where Singapore is a \violator with Hoiho and find agreement for 59 addresses (67\%), indicating that Singapore is actually a \violator and not likely to be a geolocation error.
We believe that Singapore actually makes sense because both countries are major data center hubs and are connected by multiple submarine cables (e.g. SEA-ME-WE 5~\cite{seamewe5}).

\subsection{Anycast}
\label{app:anycast}

As in our pilot study, we compare the IP addresses of the target websites 
against 
the Anycast Census dataset~\cite{anycast_census},
using the data collected closest in time to our traceroute collection (March 21, 2024) and find that 176 targets appear to be anycast (63.6\% 
of which are hosted by Cloudflare).
For traceroutes to these targets, we 
identify \emph{anycast-\violators}: countries on the path that are neither the source country nor any of the potential destination countries reported by the dataset. We find that 23\% (395) of the 1,685 traceroutes to anycast destinations have at least one anycast-\violator.  However, in all but one case, these anycast-\violators match the existing \violators seen in the non-anycast traceroutes for that country. There is one case where a probe from the U.A.E. supposedly directly connects to hops in Thailand before reaching the destination IP address, which the Anycast Census reports is hosted in Australia, Japan, India, Indonesia, or South Korea.  Unfortunately,  latency analysis is inconclusive given the close geographic proximity of Thailand to several of the potential host countries and we are unable to corroborate the purported Thai hops with other data sources such as Hoiho~\cite{hoihopaper} so are uncertain whether it represents an additional \violator or not;
otherwise the set of \violators for anycast and unicast destinations in our dataset are the same.

\relatedworktable

\section{Related Work}\label{sec:related}

Our work bridges two groups of prior work: analyzing the foreign dependencies of government websites~\cite{boeira2023traffic,houser2022comprehensive,hsiao2019investigation,jonker2022ru,rkumar:imc24,singanamalla2020accept,sommese2022assessing}, and analyzing \violators~\cite{candela2021worldwide,edmundson2018nation,fanou2015diversity,gueye2018prevalence,gupta2014peering,shah2016towards}. To the best of our knowledge, our study is the first to develop a method to identify and validate \violators, and apply it to government websites. 

\subsection{Government Website Dependencies}

Our work complements the growing body of literature that analyzed foreign dependencies of government websites' DNS infrastructure~\cite{boeira2023traffic,houser2022comprehensive,sommese2022assessing}, certificate authorities~\cite{hsiao2019investigation,singanamalla2020accept}, hosting providers \cite{boeira2023traffic,jonker2022ru,rkumar:imc24}, and content providers~\cite{hsiao2019investigation}.
One study of hosting providers~\cite{rkumar:imc24} uses a methodology similar to ours.
A 2022 study finds Russian domains are almost entirely convergent~\cite{jonker2022ru}, similarly to a 2023 study that finds some government websites for Brazil, India, and South Africa are mostly convergent~\cite{boeira2023traffic}. Finally, a 2019 study 
found that all G7 countries have some government websites that depend on foreign content providers~\cite{hsiao2019investigation}.


\subsection{Identifying \Violators}

While we are not the first to attempt to measure the prevalence of \violators (of which convergent paths with \violators are often called ``tromboning'' or ``boomerang'' paths), most prior studies are more general in the paths they consider and lack rigorous validation.
We consider whether prior work~\cite{candela2021worldwide,edmundson2018nation,fanou2015diversity,gueye2018prevalence,gupta2014peering,shah2016towards} addresses four of the five challenges we identify (the challenge of validating government websites is not applicable): validating geolocation of 1) hops and 2) vantage points, considering 3) traceroute reachability and unreliability, and 4) anycast destinations.
\ifshort
\else
We summarize each paper's ability to navigate these challenges in Table~\ref{tab:related-work}.
\fi

Only one of these studies attempts to validate the geolocation of intermediate hops: Fanou \emph{et al.}~\cite{fanou2015diversity} cross-validates multiple geolocation datasets with each other and with ping latencies. Their study, as well as one by Gupta \emph{et al.}~\cite{gupta2014peering}, both address the challenge of validating geolocation of vantage points by controlling the deployment of these vantage points. 
Fanou \emph{et al.} and Shah \emph{et al.}~\cite{shah2016towards} consider the unreliability of traceroute responses; the former notes the impact of unlabeled hops on traceroute latencies but does not separate or discard these traceroutes, while the latter discards traceroutes with fewer than three hops responding. 
Only Candela \emph{et al.}~\cite{candela2021worldwide} considers traceroute reachability, by discarding all traceroutes that do not reach the destination IP address; however, we believe this may bias results (Section~\ref{sec:highlevel:reachability}). Finally, none of these studies explain how anycast destinations complicate their definitions of boomerang routes or detours. While our work does not solve all of these challenges, to the best of our knowledge, it is the first to consider the impact of each of these challenges on our findings.

\section{Recommendations}\label{sec:conclusion}


Traffic sovereignty has become increasingly important for governments that wish to understand how their network traffic flows through other countries. However, it is extremely difficult to understand these dependencies for a large number of countries due to five core challenges: IP geolocation, collecting government websites, finding appropriate vantage points, interpreting traceroute responses, and handling anycast websites.
Our work approaches these challenges through corroboration with external sources and extensive manual validation with some success, but we do not solve all of these challenges. As the research community continues to tackle these challenges, our framework will also improve.

Our work leads to several recommendations for measurement researchers and opportunities for future work.
Our methodology relies on extensive manual effort (e.g. visiting government websites, identifying geolocation errors) to validate our dataset. Automating these manual efforts could enable larger-scale analyses of \violators.
We recommend that researchers use extra caution when conducting measurements from third-party vantage points, as self-reported geolocation may be incorrect (Section~\ref{sec:navigating:probes}). We also recommend that the research community continue to work on improving geolocation accuracy for the Internet core and identifying IP squatting. For identifying \violators, promising areas of future work include inferring \violators from unlabeled hops.
Future work could also investigate the presence of \violators in paths for TLS handshakes.
Finally, more targeted, country-level studies of \violators can sidestep many challenges of a global study (e.g., by deploying vantage points with known geolocations).


\section*{Acknowledgments}

We would like to thank Gautam Akiwate, Esteban Carisimo, Miro Haller, Liz Izhikevich, Ties de Kock, Dave Levin, Enze Liu, Alex Marder, Marios Mavropoulos Papoudas, Loqman Salamatian, Aaron Schulman, Yuval Shavitt, and Leila Scola for all of their feedback, data, and advice.

We also thank Christine Alvarado, Ruanqianqian (Lisa) Huang, and Javahir Abbasova for running the Early Research Scholars Program (ERSP), through which our undergraduate students joined this project.

This material is based upon work supported by the National Science Foundation
Graduate Research Fellowship Program under Grant No. DGE-2038238. Any opinions,
findings, and conclusions or recommendations expressed in this material are
those of the author(s) and do not necessarily reflect the views of the National
Science Foundation.

\bibliographystyle{plain}
\bibliography{paper}

@inproceedings {andromeda,
    author = {Michael Dalton and others},
    title = "{Andromeda: Performance, Isolation, and Velocity at Scale in Cloud Network Virtualization}",
    booktitle = {Proc. of 15th USENIX NSDI},
    year = {2018},
}

@article{macaskill13gchq,
    author = {Ewen MacAskill and Julian Borger and Nick Hopkins and Nick Davies and James Ball},
    title = "{GCHQ taps fibre-optic cables for secret access to world's communications}",
    journal = {The Guardian},
    year = {2013},
    month = {June},
    day = {21},
    howpublished = {\url{https://www.theguardian.com/uk/2013/jun/21/gchq-cables-secret-world-communications-nsa}}
}

@misc{cavpn,
    title = "{Making the right connections with VPN}",
    author = {Government of Canada},
    note = {Accessed 2025-05-15. \url{https://www.canada.ca/en/shared-services/campaigns/stories/right-connections.html}}
}

@misc{freedomhouse,
    key = {Freedom House Scores},
    title = "{Internet Freedom Scores}",
    author = {Freedom House},
    note = {Accessed 2024-10-04. \url{https://freedomhouse.org/countries/freedom-net/scores}}
}

@inproceedings{censys15,
  author    = {Zakir Durumeric and David Adrian and Ariana Mirian 
               and Michael Bailey and J. Alex Halderman},
  title     = "{A Search Engine Backed by Internet-Wide Scanning}",
  booktitle = {Proc of. 22nd {ACM} CCS},
  year      = 2015
}

@inproceedings{singanamalla2020accept,
  title="{Accept the Risk and Continue: Measuring the Long Tail of
Government https Adoption}",
  author={Singanamalla, Sudheesh and Jang, Esther Han Beol and Anderson, Richard and Kohno, Tadayoshi and Heimerl, Kurtis},
  booktitle={Proc. of 20th ACM IMC},
  year={2020}
}

@inproceedings{rkumar:imc24,
    title = "{Of Choices and Control - A Comparative Analysis of Government Hosting}",
    author = {Rashna Kumar and Esteban Carisimo and Lukas De Angelis Riva and Mauricio Buzzone and Fabián E. Bustamante and Ihsan Ayyub Qazi and Mariano G. Beiro},
    year = {2024},
    date = {2024-11-04},
    booktitle = {Proc. of 24th ACM IMC},
}

@article{izhikevich2024trust,
      title="{Trust, But Verify, Operator-Reported Geolocation}", 
      author={Katherine Izhikevich and Ben Du and Sumanth Rao and Alisha Ukani and Liz Izhikevich},
      year={2024},
      url={https://arxiv.org/abs/2409.19109}, 
      journal = {arXiv preprint arXiv:2409.19109},
}

@misc{netacuity,
  key = "{NetAcuity}",
  title = "{NetAcuity}",
  note   = {https://digitalelement.com/solutions/ip-location-targeting/netacuity}
}

@misc{ipinfo,
  title  = {{IP} Geolocation {API}},
  author = {ipinfo.io}
}

@inproceedings{luckie2016bdrmap,
  title="{bdrmap: Inference of borders between IP networks}",
  author={Luckie, Matthew and Dhamdhere, Amogh and Huffaker, Bradley and Clark, David and Claffy, kc},
  booktitle={Proc. of 16th ACM IMC},
  year={2016}
}

@inproceedings{zmap,
  title="{ZMap: Fast Internet-wide Scanning and Its Security Applications}",
  author={Durumeric, Zakir and Wustrow, Eric and Halderman, J Alex},
  booktitle={USENIX Security Symposium},
  volume={8},
  year={2013}
}

@inproceedings{zdns,
  author    = {Izhikevich, Liz and Akiwate, Gautam and Berger, Briana and Drakontaidis, Spencer and Ascheman, Anna and Pearce, Paul and Adrian, David and Durumeric, Zakir},
  booktitle = {Proc. of 22nd ACM IMC},
  title     = "{ZDNS: A Fast DNS Toolkit for Internet Measurement}",
  year      = {2022},
  series    = {IMC '22}
}

@inproceedings{houser2022comprehensive,
  title        = "{A Comprehensive, Longitudinal Study of Government DNS Deployment at Global Scale}",
  author       = {Houser, Rebekah and Hao, Shuai and Cotton, Chase and Wang, Haining},
  booktitle    = {Proc. of 52nd IEEE/IFIP DSN},
  year         = {2022},
}

@article{candela2021worldwide,
  title={A worldwide study on the geographic locality of Internet routes},
  author={Candela, Massimo and Luconi, Valerio and Vecchio, Alessio},
  journal={Computer Networks},
  year={2021},
  
}

@inproceedings{edmundson2018nation,
  title="{Nation-State Hegemony in Internet Routing}",
  author={Edmundson, Anne and Ensafi, Roya and Feamster, Nick and Rexford, Jennifer},
  booktitle={Proc. of 1st ACM COMPASS},
  year={2018}
}

@inproceedings{shah2016towards,
  title     = "{Towards Characterizing International Routing Detours}",
  author    = {Shah, Anant and Fontugne, Romain and Papadopoulos, Christos},
  booktitle = {Proc. of the 12th AINTEC},
  year      = {2016}
}

@inproceedings{obar2012internet,
  title     = "{Internet Surveillance and Boomerang Routing: A Call for Canadian Network Sovereignty}",
  author    = {Obar, Jonathan A and Clement, Andrew},
  booktitle = {Proc. of TEM 2013},
  year      = {2012}
}

@inproceedings{jonker2022ru,
  title = "{Where .ru? Assessing the Impact of Conflict
on Russian Domain Infrastructure}",
  author={Jonker, Mattijs and Akiwate, Gautam and Affinito, Antonia and Claffy, KC and Botta, Alessio and Voelker, Geoffrey M and van Rijswijk-Deij, Roland and Savage, Stefan},
  booktitle={Proc. of 22nd ACM IMC},
  year={2022}
}

@inproceedings{sommese2022assessing,
  title="{Assessing e-Government DNS Resilience}",
  author={Sommese, Raffaele and Jonker, Mattijs and van der Ham, Jeroen and Moura, Giovane CM},
  booktitle={Proc. of 18th IEEE CNSM},
  year={2022},
}

@inproceedings{sommese2020manycast2,
      author = {Sommese, Raffaele and Bertholdo, Leandro and Akiwate, Gautam and Jonker, Mattijs and van Rijswijk-Deij, Roland and Dainotti, Alberto and Claffy, KC and Sperotto, Anna},
      title = "{MAnycast2: Using Anycast to Measure Anycast}",
      year = {2020},
      booktitle = {Proc. of 20th ACM IMC},
}

@inproceedings{hsiao2019investigation,
  title="{An Investigation of Cyber Autonomy on Government Websites}",
  author={Hsiao, Hsu-Chun and Kim, Tiffany Hyun-Jin and Ku, Yu-Ming and Chang, Chun-Ming and Chen, Hung-Fang and Chen, Yu-Jen and Wang, Chun-Wen and Jeng, Wei},
  booktitle={Proc. of ACM Web Conf.},
  year={2019}
}

@article{boeira2023traffic,
  title="{Traffic Centralization and Digital Sovereignty: An Analysis Under the Lens of DNS Servers}",
  author={Boeira, Dem{\'e}trio F and Scheid, Eder J and Franco, Muriel F and Zembruzki, Luciano and Granville, Lisandro Z},
  journal={arXiv preprint arXiv:2307.01300},
  year={2023}
}

@misc{unportal,
  author = {United Nations},
  title = {E-government knowledgebase country data},
  note   = {https://publicadministration.un.org/egovkb/en-us/Resources/Country-URLs}
}

@misc{un_nsgt,
  author       = {United Nations},
  note = {https://www.un.org/dppa/decolonization/en/nsgt},
  title        = {Non-Self-Governing Territories},
  year         = {2022}
}

@article{satariano22russia,
  author          = {Satariano, Adam and Reinhard, Scott},
  journal         = {The New York Times},
  title           = {How Russia Took Over Ukraine's Internet in Occupied Territories},
  year = {2022},
  month = {08},
  day = {09},
  urldate         = {2023-05-17}
}

@misc{pfx2as,
    title={{Routeviews Prefix to AS mappings Dataset (pfx2as) for IPv4 and IPv6}},
    author={CAIDA},
    url={https://www.caida.org/catalog/datasets/routeviews-prefix2as},
    year={2022}
}

@inproceedings{luckie2021learning,
  title="{Learning to Extract Geographic Information from Internet Router Hostnames}",
  author={Luckie, Matthew and Huffaker, Bradley and Marder, Alexander and Bischof, Zachary and Fletcher, Marianne and Claffy, KC},
  booktitle={Proc. of 17th CoNEXT},
  year={2021}
}

@article{brazil,
  author          = {Israel, Esteban and Boadle, Anthony},
  journal         = {Reuters},
  title           = "{Brazil conference will plot Internet's future post NSA spying}",
  year            = {2014},
  month           = {April},
  day             = {22},
  note    = {https://www.reuters.com/article/us-internet-conference-idUSBREA3L1OJ20140422}
}

@inproceedings{Geo-database-comparison,
 author = {Gharaibeh, Manaf and Shah, Anant and Huffaker, Bradley and Zhang, Han and Ensafi, Roya and Papadopoulos, Christos},
 title = "{A Look at Router Geolocation in Public and Commercial Databases}",
 booktitle = {Proc. of 17th ACM IMC},
 year = {2017},
}

@article{ipmap,
    author = {Du, Ben and Candela, Massimo and Huffaker, Bradley and Snoeren, Alex C. and Claffy, KC},
    title = "{RIPE IPmap Active Geolocation: Mechanism and Performance Evaluation}",
    year = {2020},
    issue_date = {April 2020},
    journal = {ACM SIGCOMM Comput. Commun. Review},
    month = {May},
}

@inproceedings{hoihopaper,
    author = {Luckie, Matthew and Huffaker, Bradley and Claffy, KC},
    title = "{Learning Regexes to Extract Router Names from Hostnames}",
    year = {2019},
    booktitle = {Proc. of 19th ACM IMC},
}

@inproceedings{fontugne2017pinpointing,
  title="{Pinpointing Delay and Forwarding Anomalies Using Large-Scale Traceroute Measurements}",
  author={Fontugne, Romain and Pelsser, Cristel and Aben, Emile and Bush, Randy},
  booktitle={Proc. of 17th ACM IMC},
  year={2017}
}

@article{anycast_census,
    title="{MAnycast Reloaded: a Tool for an Open, Fast, Responsible and Efficient Daily Anycast Census}", 
    author={Remi Hendriks and Matthew Luckie and Mattijs Jonker and Raffaele Sommese and Roland van Rijswijk-Deij},
    year={2025},
    journal = {arXiv preprint arXiv:2503.20554},
}

@book{underseanetwork,
    author = {Starosielski, Nicole},
    title = "{The Undersea Network}",
    publisher = {Duke University Press},
    year = {2015}
}

@article{donnet2012revealing,
  title="{Revealing MPLS Tunnels Obscured from Traceroute}",
  author={Donnet, Benoit and Luckie, Matthew and M{\'e}rindol, Pascal and Pansiot, Jean-Jacques},
  journal={ACM SIGCOMM Comput. Commun. Review},
  year={2012},
}

@article{salamatian2023squats,
  title="{Who Squats IPv4 Addresses?}",
  author={Salamatian, Loqman and Arnold, Todd and Cunha, {\'I}talo and Zhu, Jiangchen and Zhang, Yunfan and Katz-Bassett, Ethan and Calder, Matt},
  journal={ACM SIGCOMM Comput. Commun. Review},
  year={2023}, 
}

@misc{seamewe5,
    title = "{SEA-ME-WE 5 Maps}",
    key = {SEAMEWE},
    url = {\url{https://seamewe5.com/route/swm5-maps/}}
}

@misc{dot_gov_domains,
  author = {U.S. Cybersecurity \& Infrastructure Security Agency},
  note    = {\url{https://github.com/cisagov/dotgov-data}},
  title  = {.gov Data}
}

@misc{military_domains,
  author = {U.S. Department of Defense},
  note    = {\url{https://www.defense.gov/Resources/Military-Departments/DOD-Websites/}},
  title  = {DOD Websites}
}

@misc{belarus_domains,
  author       = {Ministry of Foreign Affairs of the Republic of Belarus},
  note = {https://mfa.gov.by/en/links/},
  title        = {Links to Official Belarusian Web-Sites},
}

@misc{brazil_domains,
  author = {Brazil Ministry of Management and Innovation in Public Services},
  howpublished = {\url{https://dados.gov.br/dados/conjuntos-dados}/dados-da-estrutura-organizacional-do-poder-executivo-federal-sistema-siorg},
  title = {Data from the Organizational Structure of the Federal Executive Branch (SIORG System)}
}

@misc{canada_domains,
  author = {Government of Canada},
  howpublished = {\url{https://www.canada.ca/en/government/dept.html}},
  title = {Departments and agencies}
}

@misc{philippines_domains_official,
  author       = {Republic of the Philippines},
  howpublished = {\url{https://www.officialgazette.gov.ph/lists}/government-websites/},
  title        = {List of Government Websites},
}

@misc{philippines_domains_embassy,
  author       = {The Philippine Embassy of Wellington, New Zealand},
  howpublished = {\url{https://www.officialgazette.gov.ph/lists}/government-websites/},
  title        = {PH Government Websites},
}

@misc{singapore_domains,
  author       = {Government of Singapore},
  howpublished = {\url{https://www.gov.sg/trusted-sites}},
  title        = {Trusted Sites},
}

@misc{taiwan_domains,
  author = {Ministry of Foreign Affairs, Republic of China (Taiwan)},
  howpublished = {\url{https://www.taiwan.gov.tw/3866.php?xq\_xcat=5}},
  title = {Government Agencies}
}

@misc{israel_domains_ministries,
  author = {Government of Israel},
  howpublished = {\url{https://www.gov.il/en/departments}},
  title = {Government Ministries of Israel}
}

@misc{israel_domains_gov_orgs,
  author = {Government Procurement Administration of Israel},
  howpublished = {\url{https://mr.gov.il/ilgstorefront/en/gov-websites}},
  title = {Governmental and Public Organizations}
}

@misc{china_wiki,
  author = {Wikipedia},
  howpublished = {\url{https://en.wikipedia.org/wiki/Category:Government_agencies_of_China}},
  title = {Category:Government agencies of China}
}

@misc{kenyapres,
  author = {Office of the President},
  howpublished = {\url{https://www.president.go.ke/ministries-ke/}},
  title = {Category:Government ministries of Kenya}
}

@misc{kenyadps,
  author = {Office of the Deputy President},
  howpublished = {\url{https://www.devolution.go.ke/county-websites/}},
  title = {Category:Government ministries of Kenya}
}

@misc{uaecabinet,
  author = {UAE Cabinet},
  howpublished = {\url{https://uaecabinet.ae/en/ministries-and-federal-authorities}},
  title = {Category:Government ministries of the UAE}
}

@misc{maxmind,
 title="{MaxMind: IP Geolocation and Online Fraud Prevention}",
 key = {MaxMind},
 url={https://www.maxmind.com},
 year={2024}
}

@inproceedings{replicationgeo,
    author = {Darwich, Omar and Rimlinger, Hugo and Dreyfus, Milo and Gouel, Matthieu and Vermeulen, Kevin},
    title = "{Replication: Towards a Publicly Available Internet Scale IP Geolocation Dataset}",
    year = {2023},
    booktitle = {Proc. of 23rd ACM IMC},
}

@inproceedings{gouel2021ip,
  title="{IP Geolocation Database Stability and Implications for Network Research}",
  author={Gouel, Matthieu and Vermeulen, Kevin and Fourmaux, Olivier and Friedman, Timur and Beverly, Robert},
  booktitle={Proc. of IEEE/IFIP TMA},
  year={2021}
}

@misc{ntt-network-map,
    author = {NTT America},
    title = "{Network Map}",
    url = {\url{https://services.global.ntt/-/media/ntt/global/services-and-products/networks/global-ip-network/landing-page/north-america-network-map-2023.png?rev=f170cc67d9214415a2afd7922f5720d8}},
    urldate = {2024-05-01}
}

@misc{belarus-net-freedom,
    author = {Freedom House},
    title = "{Belarus: Freedom on the Net 2022}",
    url = {\url{https://freedomhouse.org/country/belarus/freedom-net/2023}},
    urldate = {2024-05-05},
    year = {2022}
}

@misc{belarus-amnesty,
    author = {Amnesty International},
    title = "{Belarus uses telecoms firms to stifle dissent}",
    url = {\url{https://www.amnesty.org/en/latest/press-release/2016/07/belarus-uses-telecoms-firms-to-stifle-dissent/}},
    urldate = {2024-05-05},
    year = {2016},
    month = {July},
    day = {7}
}

@inproceedings{2014-luckie-slda,
  author = {Luckie, Matthew and Claffy, KC},
  title = "{A Second Look at Detecting Third-Party Addresses in Traceroute Traces with the IP Timestamp Option}",
  year = {2014},
  booktitle = {In Proc. of PAM}
}

@misc{ris,
    author={RIPE NCC},
    title="{Routing Information System (RIS)}",
    url="\url{https://www.ripe.net/analyse/internet-measurements/routing-information-service-ris}",
    year={2024}
}

@misc{routeviews,
    author={University of Oregon},
    title="{Route Views Project}",
    url="\url{http://www.routeviews.org/routeviews}",
    year={2024}
}

@inproceedings{gueye2018prevalence,
  title="{On the Prevalence of Boomerang Routing in Africa: Analysis and Potential Solutions}",
  author={Gueye, Assane and Mbaye, Babacar},
  booktitle={Proc. of InterSol},
  year={2018},
}

@inproceedings{gupta2014peering,
  title="{Peering at the Internet's Frontier: A First Look at ISP Interconnectivity in Africa}",
  author={Gupta, Arpit and Calder, Matt and Feamster, Nick and Chetty, Marshini and Calandro, Enrico and Katz-Bassett, Ethan},
  booktitle={Proc. of 15th PAM},
  year={2014}
}

@inproceedings{fanou2015diversity,
  title="{On the Diversity of Interdomain Routing in Africa}",
  author={Fanou, Rod{\'e}rick and Francois, Pierre and Aben, Emile},
  booktitle={Proc. of 16th PAM},
  year={2015},
}

\appendix
\section{Ethics}

We attempted to reduce the burden of our traceroutes to the domains we probe by using at most 10 RIPE Atlas probes.
\section{Collection Process for Official Government Websites}
\label{appendix:gov_website_collection}

In this section we describe how we collected official government websites for
each country in our focused study:



\textbf{Belarus}. The Belarus Ministry of Foreign Affairs publishes a list of
145 official government websites~\cite{belarus_domains}. We visited this website on January 30, 2024;
it is unknown when the website was last updated.

\textbf{Brazil}. The Brazilian Ministry of Management and Innovation in Public
Services publishes a dataset called ``Organizational Structure of the Federal
Executive Branch''~\cite{brazil_domains}. The dataset is updated monthly and we
used the latest available dataset, which was created on January 2, 2024. The
dataset contains 1,494 unique URLs.

\textbf{Canada}. The Canadian government publishes a website with its
departments and agencies~\cite{canada_domains}, which was last updated on
October 4, 2023. We scraped the links on this website on January 24, 2024. Some
of these links were to internal pages; in this case, we first visited the
internal link and checked if it redirected to an external website. If not, we
collected all internal links in a \texttt{main} HTML element. This process
resulted in 138 unique URLs.

\textbf{China}. 
We used Wikipedia to find a list of government agencies belonging to China,
with urls linking to another Wikipedia page about each agency~\cite{china_wiki}. 
On January 23, 2024 we manually opened each Wikipedia page and looked for either an ``External Links'' 
section that linked to the ageny's own website or we found the website linked 
in the ``Agency Overview'' under the ``Website'' section. This process resulted in 
42 unique URLs.




\textbf{Israel}. The Israeli government publishes a website with its government 
ministries~\cite{israel_domains_ministries} and the Government Procurement Administration
publishes a list of government websites~\cite{israel_domains_gov_orgs}. We scraped the 
links on January 22, 2024. This process resulted in 309 unique URLs.


\textbf{Kenya}. The Kenyan Office of the President \cite{kenyapres} provides a list of all national ministries.
Their Office of the Deputy President \cite{kenyadps} provides a list of 47 county government websites. We 
scraped those 2 lists on Jan 24, 2024 and obtained 89 unique URLs.


\textbf{Philippines}. The Philippines government publishes a list of official
government websites for some of its federal
organizations~\cite{philippines_domains_official}. We supplemented this list
with a list of government websites from the Philippines Embassy of New
Zealand~\cite{philippines_domains_embassy}. We visited both websites on August
28, 2023 and collected a total of 79 unique URLs.

\textbf{Singapore}. The Singapore government claims that most of their
government websites have the form \texttt{*.gov.sg}; however, they publish a
list of 125 official government websites that do not follow this
pattern~\cite{singapore_domains}. We visited this website on January 24, 2024.
It is unknown when this website was last updated.



\textbf{Taiwan}. The Taiwanese government publishes a list of government
agencies on their website~\cite{taiwan_domains}. We visited this website on
January 26, 2024 and collected a total of 295 unique URLs.

\textbf{United Arab Emirates}. The Cabinet of the UAE provides a list of ministries
and federal authorities \cite{uaecabinet}. We scraped the website on Jan 24, 2024
and obtained 46 unique URLs.


\textbf{USA}. As mentioned in Section~\ref{sec:method}, we combine three
official data sources. The first source is a set of URLs ending in
\texttt{.gov}, which is updated every day~\cite{dot_gov_domains} and
contains 9,833 unique URLs. The second source is a set of non-\texttt{.gov}
domains; we use the \texttt{1\_govt\_urls\_full.csv} file, which contains
domains across federal, state, regional, county, and local levels, as well
as native sovereign nations (i.e., tribal nations) and quasigovernmental
organizations. This file contains 9,226 unique URLs. The third and final
source is a set of U.S. military domains from the Department of Defense
website~\cite{military_domains}, from which we exclude social media websites
for a total of 593 unique URLs.

In total, these data sources contained 19,299 unique URLs. This number is
more than the sum of the URLs in each data source as there were some
duplicates between the military URLs and the other two datasets.

Each of these sources were visited on January 30, 2024; at the time we
visited them, the \texttt{.gov} dataset was last updated that same day, and
the non-\texttt{.gov} domains were last updated on January 30, 2023 (a year
prior). It is unknown when the website with military domains was last
updated.

\section{Additional Validation Details}
\label{app:valdetail}

We use Censys~\cite{censys15} to expand our set of government domains collected from authoritative sources (Section~\ref{sec:method:censys}).
This process identifies a large number of URLs in each country, but some of these may be unreachable, not serve webpages, or may not belong to the government. We considered filtering to only ``trusted certificates,'' i.e. certificates where the certificate chain can be followed to a root certificate found in a major root trust store.
However, using only trusted certificates filters out legitimate government websites. For example, 23 legitimate Belarussian governmental websites (ending in \texttt{.gov.by}) use untrusted TLS certificates.
Instead, 
we filter the Censys results to subdomains of the domains and eTLD+1's that we obtained through authoritative sources (Section~\ref{sec:method:censys}) and manually inspect each website.


\subsection{Impact of Validation}

\heatmapconvergentviolators


In Figure~\ref{fig:heatmap_convergent_violators} we present a heatmap of the paths to convergent websites that contain \violators, for both the full and the refined datasets (see Figure~\ref{fig:heatmap_violators} for a similar heatmap for both convergent and divergent destinations). After data sanitization, all but one \violator disappear for China, as most of the \violators were experienced by the probes we exclude.

We find relatively few geolocation errors overall because 1) we use a high threshold of speed-of-light violation for identifying errors, and 2) we focus our efforts on confirming \violators identified by the geolocation dataset we use;
we do not try to find additional \violators that were not identified by IPinfo.

\subsection{Interesting anecdotes}
\label{app:int}

This subsection contains a variety of potentially interesting \violators we uncovered during our manual validation.

\subsubsection{\Violator Path due to Telxius Cables}

We find two traceroutes to the destination website \texttt{csjt.jus.br} (for Brazil's Superior Council of Labor Justice) that transit Spain and the U.S. en route to the destination website in Colombia (hosted by Amazon). Both paths traverse Telef\'{o}nica Global Solutions, and within this network the paths are first routed to the U.S. and then Spain; after these hops, we see four unlabeled responses before reaching a Colombia IP address associated with Amazon.

Based on the publicly available network maps from Telxius (a child company of Telef\'{o}nica), we see that this path maps onto high-capacity submarine cables: the Firminia or Brusa cables that connect Brazil to the U.S., and the Marea cable that connects the U.S. to Spain~\cite{ntt-network-map}. However, this map does not show a direct connection from Spain to Colombia. If the path were to continue within the Telxius network and the public network map, the path would need to go back to the U.S., go from Virginia Beach to Ashburn via terrestrial backhaul, go to Jacksonville via an extended fiber route, and then go to Colombia via a submarine cable.
Interestingly, the Telxius SAm-1 cable directly connects Brazil and Colombia and thus offers a path without \violators; however, we do not see this route (or any other IP addresses for Colombia) in our data.

Hoiho does not contain any validation information for this traceroute.
Latencies spike from around 7ms to 130ms for the US hop and then stay around 130ms for the hops to Spain and Colombia.

\subsubsection{US as \violator for China}

We find two traceroutes from a single probe (53274) that have the U.S. as a \violator. For this probe’s paths without \violators, the third hop is always an IP address announced by No.31,Jin-rong Street (AS4134). However, in the paths with \violators, hops 3 and 4 remain in the 192.0.0.0/8 space, and the fifth hop geolocates to the U.S. and is announced by EGIHosting (AS18779). The latency for the fifth hop indicates that visiting the U.S. is plausible, but we are unable to further confirm this behavior.

\subsubsection{Canada as \violator for US}

A Seattle probe produces one path with one \violator hop, which geolocates to Vancouver. The IP address, as well as the ones immediately before and after in the path, are within the AT\&T network. We are unable to corroborate the geolocation, as Hoiho does not have geolocation information for any of these AT\&T IP addresses. However, IPinfo indicates that the hops preceding the \violator remain in Seattle and Portland, indicating that this path is plausible.

\subsubsection{\violators for Canada}

The U.S. is the most prevalent \violator for Canada. In particular, we find a common pattern for the 168 paths with \violators to convergent targets: the source RIPE Atlas probes are located in Vancouver, and the destination is hosted in cities in eastern Canada such as Ottowa and Montreal. The \violator hops traverse mostly northern U.S. cities such as Seattle, Chicago, New York, but also some southern cities like Virginia and Dallas. The \violator hops are announced by ASNs for major U.S. cloud providers like Microsoft and Amazon, as well as Tier-1 providers like Cogent and Arelion. 

The divergent paths \textit{without} \violators all follow a similar pattern: once the traceroutes enter the U.S., they stay inside the U.S. until reaching their destinations.
The divergent paths \textit{with} \violators have destinations hosted in two countries: Sweden (Optimizely AB, AS30811) and Ireland (Amazon, AS16509). These paths fall into the following two patterns: (1) the traffic is first routed by the Tier-1 transit provider Arelion into the U.S. and then to Denmark and Sweden, and (2) the traffic is routed by Cogent directly from Canada to the U.K. and finally to Ireland. The divergent paths with \violator only include the U.S. and U.K. before reaching their destination in Ireland or Sweden.
\end{document}